\documentclass[twocolumn,superscriptaddress,reprint,aps,prb]{revtex4-2}
\usepackage{amssymb}
\usepackage{amsmath}
\usepackage{graphicx}
\usepackage{dcolumn}
\usepackage{bm}
\usepackage{float}
\usepackage{xcolor}
\usepackage[normalem]{ulem}
\usepackage{titlesec}

\titleformat{\paragraph}
{\normalfont\normalsize\bfseries}{\theparagraph}{1em}{}
\titlespacing*{\paragraph}
{0pt}{3.25ex plus 1ex minus .2ex}{1.5ex plus .2ex}

\begin{document}

\title{Magnetism of kagome metals $\left(\text{Fe}_{1-x} \text{Co}_{x}\right) \text{Sn}$ studied by $\mu$SR}

\author{Yipeng Cai}
\email{Contact author: caidia52@gmail.com}
\affiliation{Department of Physics, Columbia University, New York, New York 10027, USA}
\address{Quantum Matter Institute, University of British Columbia, Vancouver, British Columbia V6T 1Z4, Canada}
\address{TRIUMF, Vancouver, British Columbia V6T 2A3, Canada}

\author{Sungwon Yoon}
\address{TRIUMF, Vancouver, British Columbia V6T 2A3, Canada}
\address{Department of Physics, Sungkyunkwan University, Suwon 16419, Republic of Korea}

\author{Qi Sheng}
\address{Department of Physics, Columbia University, New York, New York 10027, USA}

\author{Guoqiang Zhao}
\address{Department of Physics, Columbia University, New York, New York 10027, USA}
\address{Institute of Physics, Chinese Academy of Sciences, Beijing 100190, China}

\author{Eric Francis Seewald}
\author{Sanat Ghosh}
\author{Julian Ingham}
\author{Abhay Narayan Pasupathy}
\author{Raquel Queiroz}
\address{Department of Physics, Columbia University, New York, New York 10027, USA}

\author{Hechang Lei}
\address{Department of Physics and Beijing Key Laboratory of Opto-Electronic Functional Materials \& Micro-Nano Devices, Renmin University of China, Beijing, China}

\author{Yaofeng Xie}
\address{Department of Physics and Astronomy, Rice University, Houston, TX, USA}
\author{Pengcheng Dai}
\address{Department of Physics and Astronomy, Rice University, Houston, TX, USA}
\address{Smalley-Curl Institute, Rice University, Houston, TX, USA}

\author{Takashi Ito}
\address{Advanced Science Research Centre, Japan Atomic Energy Agency (JAEA), Tokai, Naka, Ibaraki 319-1195, Japan}

\author{Ruyi Ke}
\author{Robert J. Cava}
\address{Department of Chemistry, Princeton University, Princeton, New Jersey 08540, USA}

\author{Sudarshan Sharma}
\author{Mathew Pula}
\address{Department of Physics and Astronomy, McMaster University, Hamilton, Ontario, Canada L8S 4M1}

\author{Graeme M. Luke}
\address{TRIUMF, Vancouver, British Columbia V6T 2A3, Canada}
\address{Department of Physics and Astronomy, McMaster University, Hamilton, Ontario, Canada L8S 4M1}

\author{Kenji M. Kojima}
\address{Quantum Matter Institute, University of British Columbia, Vancouver, British Columbia V6T 1Z4, Canada}
\address{TRIUMF, Vancouver, British Columbia V6T 2A3, Canada}

\author{Yasutomo J. Uemura}
\email{Contact author: yu2@columbia.edu}
\address{Department of Physics, Columbia University, New York, New York 10027, USA}

\begin{abstract}
We study the magnetic properties of the metallic kagome system $\left(\mathrm{Fe}_{1-x} \mathrm{Co}_{x}\right) \mathrm{Sn}$ by a combination of Muon Spin Relaxation ($\mu \mathrm{SR}$), magnetic susceptibility and Scanning Tunneling Microscopy (STM) measurements, in single crystal specimens with Co concentrations $\mathrm{x}=0,0.11,0.8$. In the undoped antiferromagnetic compound FeSn, we find possible signatures for a previously unidentified phase that sets in at $T^*\sim 50$ K, well beneath the Neel temperature $T_N \sim 376$ K, as indicated by a peak in the relaxation rate $1/T_1$ observed in zero field (ZF) and longitudinal field (LF) $\mu \mathrm{SR}$ measurements, with a corresponding anomaly in the ac and dc-susceptibility, and an increase in the static width $1/T_2$ in ZF measurements. No signatures of spatial symmetry breaking are found in STM down to $7$ K. Related to the location and motion of muons in FeSn, we confirmed a previous report that about 40~$\%$ of the implanted muons reside at a field-cancelling high symmetry site at T $<$ 250 K, while an onset of thermal hopping changes the site occupancy at higher temperatures. In $\mathrm{Fe}_{0.89} \mathrm{Co}_{0.11} \mathrm{Sn}$, where disorder eliminated the field-cancellation effect, all the implanted muons exhibit precession and/or relaxation in the ordered state. In $\mathrm{Fe}_{0.2} \mathrm{Co}_{0.8} \mathrm{Sn}$, we find canonical spin glass behavior with freezing temperature $T_{g} \sim 3.5 \mathrm{~K}$; the ZF and LF time spectra exhibit results similar to those observed in dilute alloy spin glasses CuMn and AuFe, with a critical behavior of $1 / T_{1}$ at $T_{g}$ and $1 / \mathrm{T}_{1}\rightarrow$ as $T \rightarrow 0$. The absence of spin dynamics at low temperatures makes a clear contrast to the spin dynamics observed by $\mu \mathrm{SR}$ in many geometrically frustrated spin systems on insulating kagome, pyrochlore, and triangular lattices. The spin glass behavior of CoSn doped with dilute Fe moments is shown to originate primarily from the randomness of doped Fe moments rather than due to geometrical frustration of the underlying lattice.

\end{abstract}

\maketitle 
\onecolumngrid 
\section{INTRODUCTION}

Geometrically frustrated magnets are excellent candidate materials for exploring exotic physics, since conventional magnetic orders are suppressed. Antiferromagnetically interacting spins in such systems -- including the triangular, kagome, and pyrochlore lattices -- produce highly degenerate ground state without magnetic long-range order, resulting in tendencies towards quantum spin liquids or spin glass freezing \cite{GREEDAN2006, Gardner2010, Reimers1991, Uemura1994, Keren1996, Keren2000, Fuyaka2003, Mendels2007}. For this reason, Mott insulating kagome materials such as herbetsmithite, volborthite, and barlowite, have been studied as candidate spin liquids \cite{Helton2007, Hiroi2001, Han2014}. 

Yet, in recent years, metallic kagome systems have become a major subfield of condensed matter physics. Their bandstructure naturally hosts Dirac points, flat bands, and van Hove singularities with an unusual sublattice structure, making kagome metals a natural setting for strong correlations and band topology to interplay \cite{wang2023quantum, yin2022topological}; indeed many studies have proposed these systems to host various unconventional correlated phases \cite{Tan2011, Sun2011, Green2010, yu2012chiral, Kiesel2013, Kiesel2012, Li2022_2D, wenger2024theory, scammell2023chiral, ingham2024theory, dong2023loop, profe2024kagome}. For this reason, in the last ten years experiments have explored materials comprised of stacks of metallic kagome layers in search of evidence for interesting electronic phase transitions, the recent history of which begins with the discovery of half Heusler compounds Mn$_3$Sn and Mn$_3$Ge -- antiferromagnets exhibiting a magneto-optical Kerr effect and large anomalous Hall effect despite zero net magnetisation \cite{Nayak2016large, Kiyohara2016giant, Nakatsuji2015large, Kubler2014non} -- and of Co$_3$Sn$_2$S$_2$ soon thereafter -- a Weyl semi metal hosting a large anomalous Hall effect and surface nematic order \cite{Liu2018giant, Liu2019magnetic, Guin2019zero, nag2024pomeranchuk}. Yet these materials are relatively three-dimensional, making the role of the two-dimensional kagome motif in their interesting properties somewhat unclear.

In 2018, intermetallic compound Fe$_3$Sn$_2$ was discovered with a similar structure to Co$_3$Sn$_2$S$_2$ \cite{Ye2018,Ye2019haas}. Compared to the 3:1 intermetallics, these materials are more two-dimensional, and exhibit some aspects of the two-dimensional kagome physics. Experiment sees evidence of massive Dirac fermions gapped out by ferromagnetism with an associated large intrinsic Hall conductivity \cite{Ye2018,fang2022ferromagnetic}. The Dirac pockets can be manipulated by magnetic fields \cite{Ye2019haas}, along with skyrmion bubbles \cite{hou2017observation,du2020room,kong2023observation} and a nematic response to external magnetic fields \cite{yin2018giant}. Building on this work, the 1:1 intermetallics FeSn \cite{kang2020dirac} and CoSn \cite{Kang2020topological} were found to exhibit significantly less three-dimensionality compared to the 3:1 and 3:2 structures. ARPES reports the appearance of a relatively flat band whose dispersion is suppressed along both the in- and out-of-plane directions, exhibiting a bandwidth of $\sim 150$ meV roughly $300$ meV beneath the Fermi level in CoSn \cite{Kang2020topological}, although inelastic neutron scattering experiments were unable to identify the expected flat spin excitations from electron-hole excitations of electronic flat bands \cite{Xie2021}. The relatively flat band exists above the Fermi level in the case of FeSn; interpolating between the two compounds with Fe$_x$Co$_{1-x}$Sn for $0<x<1$ produces an interesting phase diagram hosting spin glass behaviour and non trivial antiferromagnetic order \cite{sales2019electronic,sales2021tuning}. Tunneling experiments on FeSn have seen evidence of a relatively flat band existing in the surface bandstructure, even though the kagome flat band is further away in the bulk \cite{Han2021evidence}, coexisting with Weyl fermionic surface states \cite{lin2020dirac}, and recent work has seen interesting nematic charge patterns in STM \cite{zhang2023visualizing}.

In parallel, an immense amount of recent attention has also been devoted to the 166 family RM$_6$X$_6$, including the frustrated charge density wave materials ScV$_6$Sn$_6$ and LuNb$_6$Sn$_6$ \cite{arachchige2022charge, di2023flat, korshunov2023softening, pokharel2023frustrated, cao2023competing, Huang2023, lee2024nature, jiang2024van, ortiz2024stability}, as well as the unconventional superconductors of the 135 family AV$_3$Sb$_5$ (A=K,Rb,Cs), which exhibit nematic order alongside signatures of time-reversal symmetry breaking and multiple distinct density wave orders \cite{Ortiz2020, Ortiz2021b, Xu2022, Jiang2021, mielke2022time, Chen2021roton, Liang2021, li2022rotation, Nie2022charge, liu2024absence, Guo2022switchable, asaba2024evidence, Guo2024correlated, hossain2025field, wilson2024kagome, jiang2023kagome}. A unifying theme in all these systems is the need to understand the nature of electronic symmetry breaking in the kagome lattice geometry, and its relation to unconventional superconductivity or magnetism.

\begin{figure}[H]
\centering\includegraphics[width=0.5\columnwidth]{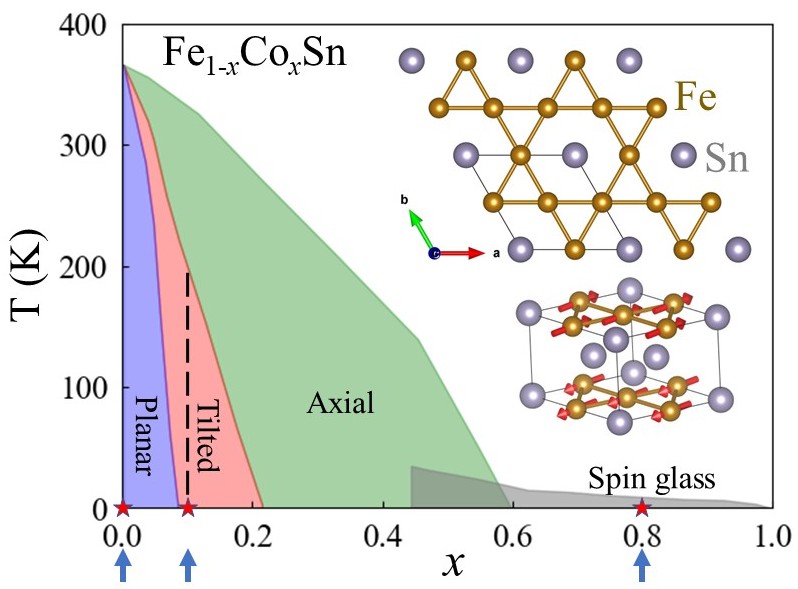}
\caption{Magnetic phase diagram of (Fe, Co)Sn. Red stars with blue arrow marks show the corresponding Co concentrations in our studies. Inset shows the kagome layer in the FeSn crystal structure and its magnetic structure below T$_N$.}
\label{Phasediagram}
\end{figure}

To further understand how electronic and magnetic orders evolve in kagome metals and the underlying mechanisms behind correlated phases, we investigated the kagome metal FeSn, shown in Fig. \ref{Phasediagram}, and explore how its magnetic properties are turned by Co substitution. FeSn features a layered kagome structure in which Fe moments align ferromagnetically within each plane and alternate between planes, forming an A-type antiferromagnetic order below the Neel temperature $\mathrm{T}_{N} \sim 376$ K. Although FeSn shares the same crystal structure and high temperature magnetic ordering as FeGe, it notably lacks the CDW formation as observed in the latter \cite{Teng2022}, raising questions about the role of electron/magnetic correlations in these materials. To probe this further, we study the solid solution series $\left(\mathrm{Fe}_{1-x} \mathrm{Co}_{x}\right) \mathrm{Sn}$, which forms a continuous kagome lattice with tunable magnetic behavior. Work by Sales et al. \cite{Sales2021} mapped out a rich phase diagram Fig. \ref{Phasediagram}, showing a rapid suppression of magnetic order with increasing Co concentration and a reorientation of the ordered moments from in-plane to out-of-plane. Near the percolation threshold at $x \sim 0.5$, the system transitions to a spin glass like regime, with freezing temperatures $T_{g}$ scale linearly with Fe concentration $x$. These observations motivate our study of how Co doping modulates the competing interactions in FeSn, with the aim of understanding the emergence of spin glass behavior in kagome metals.

To study details of magnetic properties, Muon Spin Relaxation ( $\mu \mathrm{SR}$ ) \cite{Muon_Spectroscopy} is a powerful and unique probe, which can detect: (a) the volume fraction of magnetically ordered regions in a phase-separated system; (b) static spin freezing of uniform or random spin configurations even with a very small ordered moment size; and (c) dynamic spin fluctuations in a wide fluctuation time window. A pioneering $\mu \mathrm{SR}$ study by Hartmann and Wappling in the 1980s reported that in FeSn, approximately $60 \%$ of implanted muons experience a  local field strength of $\sim 0.21 \mathrm{~T}$, while the remaining $40 \%$ appears to reside at a field cancellation site below $250$ K~\cite{Hartmann_1987}. We reexamined and performed detailed $\mu \mathrm{SR}$ measurements to further address the magnetic properties of FeSn and its evolution with Co doping. In this work, we report studies of Fe$_{1-x}$Co$_x$Sn with three concentrations $x=$ 0, 0.11, and 0.8, which exhibit, respectively: a planar magnetic ordered state, a complex reorientation ordered moment state, and a spin glass state at low temperature \cite{Meier2019, Sales2021}. 

The present work contributes to two different research fronts: (a) providing new information on the materials properties of (Fe,Co)Sn by $\mu$SR, magnetic susceptibility and STM; and (b) adding new information on the location and motion of muons in the material and to advance the methodology of the $\mu$SR technique. In the first front (a), in Sec.~\ref{Analyses} and \ref{LF-muSR}, we report a discovery of an anomaly at T $\sim$ 50~K in $1/T_1$ and $1/T_2$ measurements of $\mu$SR (in Fig.~\ref{T1}) in undoped FeSn, corresponding data from magnetic susceptibility (in Fig.~\ref{res_sus}), and STM studies (in Fig.~\ref{STM} and \ref{dI-dV}). These results could be considered as possible signatures of a new phase deep in the magnetically ordered state. Sec.~\ref{Fe20section} shows the $\mu$SR results in Co-rich sample of $x = 0.8$, which exhibits a spin glass behaviour. The ZF and LF $\mu$SR measurements show spin freezing associated with the slowing down of spin fluctuations toward the freezing temperature $T_g$ $\sim$ 3.5~K, while dynamic spin fluctuations die away at low temperatures well below $T_g$. Comparisons of these results with $\mu$SR and susceptibility data from a kagome lattice spin system SrCr$_{8}$Ga$_{4}$O$_{19}$(SCGO), described in Sec.~\ref{disc_B_spin_glass}, demonstrate that (Fe,Co)Sn in the spin glass region exhibits behaviour similar to those of the canonical dilute alloy spin glasses AuFe and CuMn, without persistent spin dynamics~\cite{Uemura1985} oftern observed in insulating magnetic systems on geometrically frustrated kagome, triangular, and/or pyrochlore lattices.

Regarding the second research front(b) with muon-specific phenomena, Sec.~\ref{usrinFeSn} includes detailed $\mu$SR data in FeSn obtained in spin-rotated and non-spin-rotated configurations and the field dependence, while Sec.~\ref{sec_muonsite} introduces simulations for muon site and local magnetic field in FeSn, and the comparisons with the observed data. The $\mu$SR results in 11$\%$ Co doped Fe$_{0.89}$Co$_{0.11}$Sn, described in Sec.~\ref{sec_Fe89}, provides an example where the 'field-cancelling effect' is absent due to disorder. General readers interested solely in materials may skip these muon-specific sections, while the detailed data and analyses in these sections provide reliability to the whole results and discussions of the present work.

\section{EXPERIMENTAL DETAILS}

\begin{figure}[H]
\centering\includegraphics[width=0.5\columnwidth]{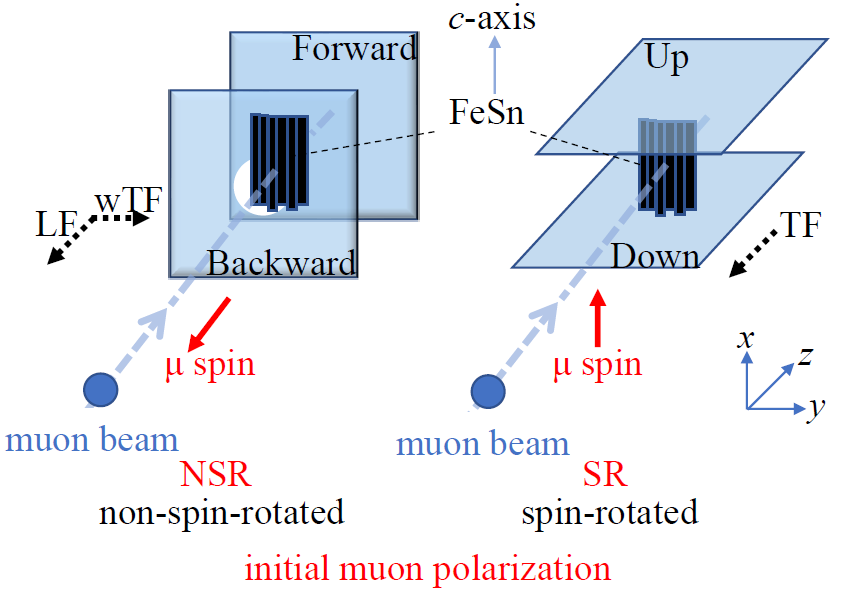}
\caption{Geometry of the $\mu$SR experiment setup at M20 beamline. Left: non spin rotated mode, where muon spin direction is parallel to the muon beam axis. Right: spin rotated mode, where spin was rotated and perpendicular to the muon beam axis. External field can be applied accordingly either parallel to the muon spin (LF) or perpendicular to the muon spin (TF). The corresponding pair of counters for collecting positrons are Forward/Backward (FB) for NSR and Up/Down (UD) for SR mode.}
\label{muSRsetup}
\end{figure}

Using methods similar to those described in Ref.~\cite{Xie2021}, high quality single crystals with $x=0$ and 0.11, having a typical size of 2$\times$2$\times$5 mm$^{3}$, were prepared at Renmin University of China, while $x = 0.8$ was prepared at Rice University. We collected magnetic susceptibility measurements from 1.9~K to 300~K on a small single crystal using a Quantum Design Magnetic Property Measurement System XL-3. We also measure the resistivity on a single crystal between 1.9 K and 300 K using a Quantum Design Physical Property Measurement System. The STM measurements were performed on FeSn crystals used in $\mu$SR studies, cleaved under ultra-high vacuum inside the STM chamber and scanned at low temperatures $\sim$ 7~K.

The $\mu$SR experiments were performed at the TRIUMF in Vancouver, Canada, using a gas flow cryostat on the M20 beamlines for measurements above 2 K on all samples, and an Oxford dilution refrigerator on the M15 beamline for measurements below 2 K on Fe$_{0.2}$Co$_{0.8}$Sn. Fig.~\ref{muSRsetup} shows the counter configuration used in our experiments at the M20 beamline. Mosaic crystals of $\sim$ 300~mg in mass for each concentration were aligned and mounted with their c-axis perpendicular to the beam direction. The a-and-b axes within the ab-plane were not aligned and remained randomly oriented. We used two different measurement modes: non-spin-rotated (NSR) and spin-rotated (SR), as illustrated in Fig.~\ref{muSRsetup}. In the SR mode, the initial muon spin direction was arranged perpendicular to the beam direction, such that the incident muon spins are parallel to the sample c-axis. $\mu$SR experiments were performed in zero applied field (ZF), longitudinal fields (LF) parallel to the initial muon spin polarization, and transverse fields (TF) perpendicular to that. All $\mu$SR data were analyzed in the time domain using the open-source muSRfit software package \cite{Suter2012}. We also employed PAW pseudopotentials with GGA-PBE from the standard density functional theory (DFT)
as implemented in QUANTUM ESPRESSO, where the muon was modelled as a proton, to simulate the muon stopping sites.

\section{RESULTS}
\subsection{$\mu$SR in FeSn} \label{usrinFeSn}
\subsubsection{Notable features in the time spectra of ZF and TF: field-cancelling site}

\begin{figure}[H]
\centering\includegraphics[width=0.8\columnwidth]{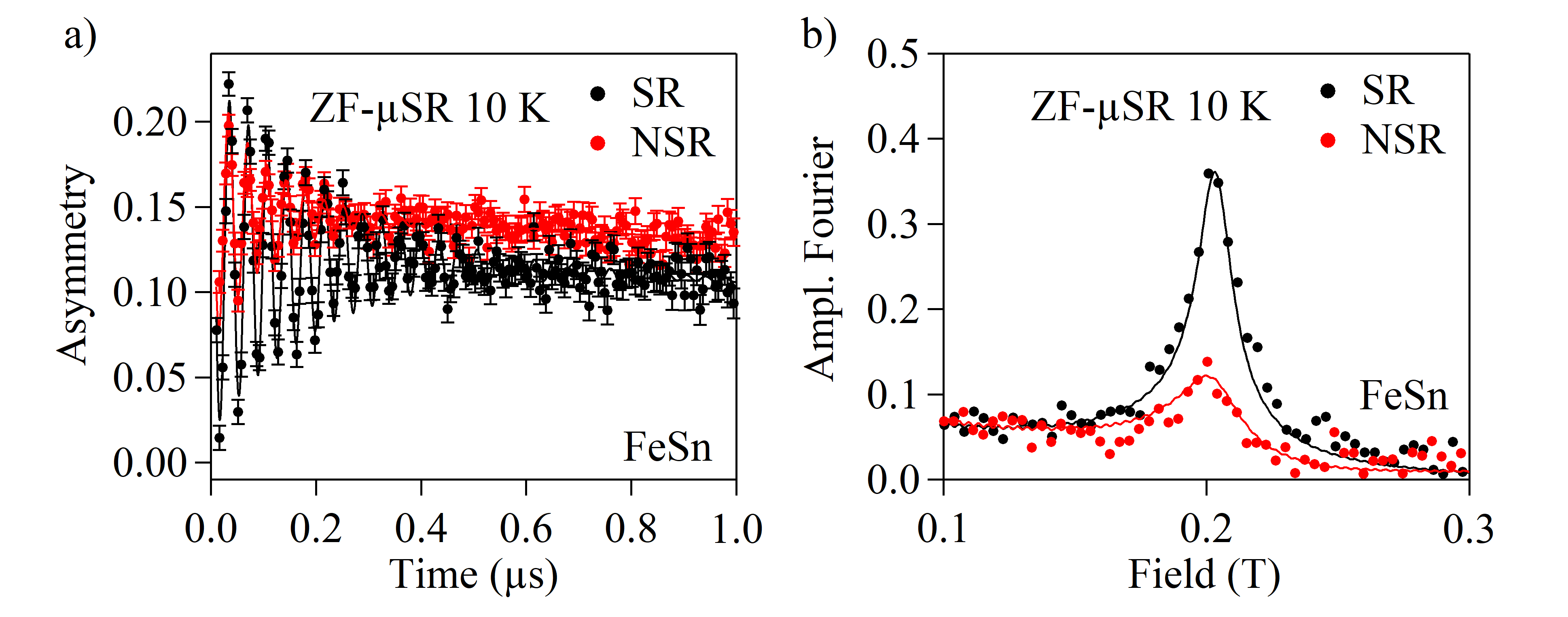}
\caption{ZF-$\mu$SR time spectra (a) and corresponding FFT spectra (b) observed in the SR and NSR modes at T = 10 K.}
\label{10K_ZFSRNSR}
\end{figure}

\begin{figure*}
\includegraphics[width=\textwidth]{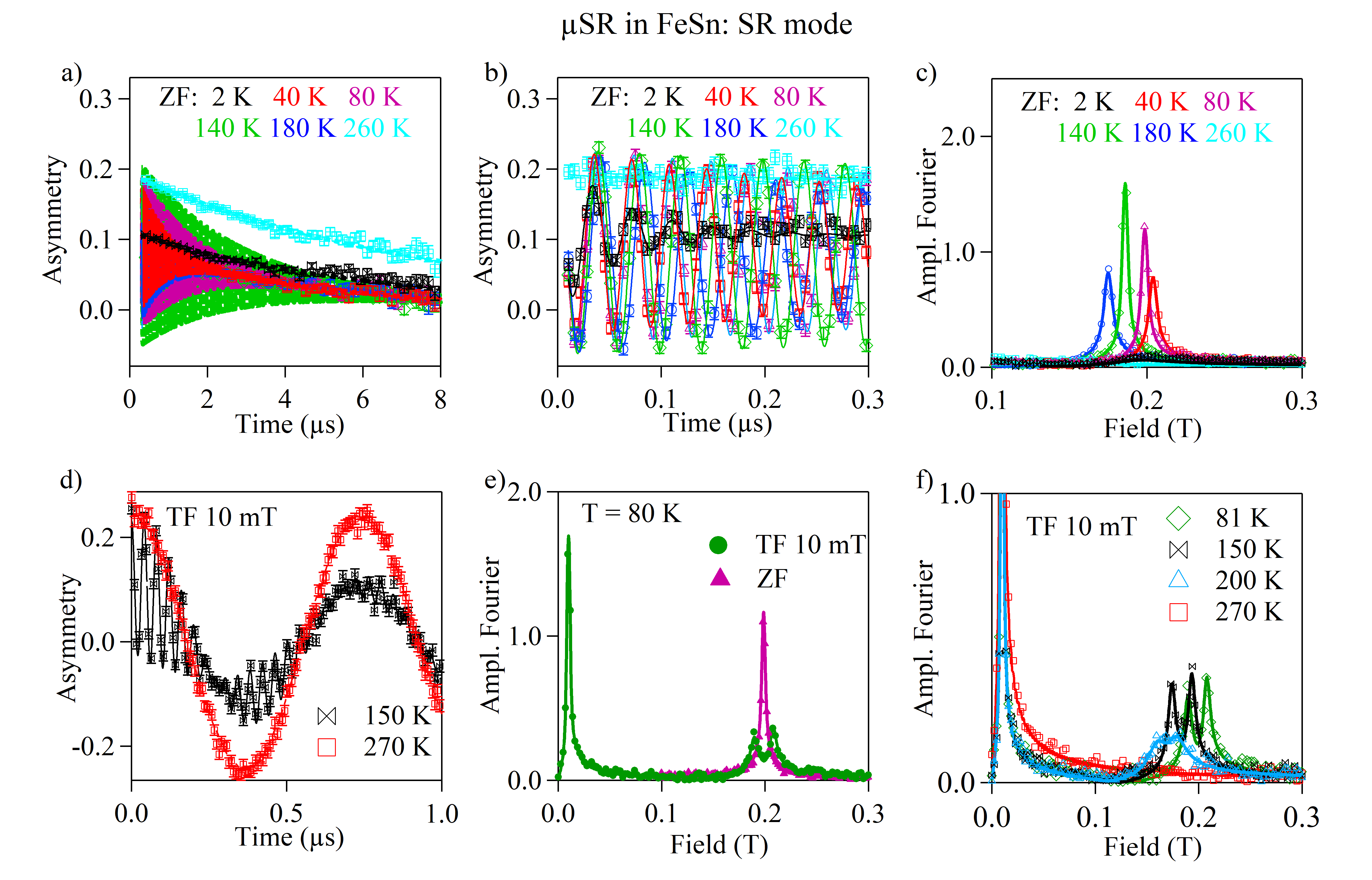}
\caption[width=\textwidth]{All lines are fitting results as described in the main text. a) ZF-SR time spectra of FeSn up to 8 $\mu s$ at selected temperatures. b) ZF-SR time spectra of FeSn within 0.3 $\mu s$. c) FFT spectra of FeSn in ZF-SR. d) wTF-SR time spectra of FeSn with an external TF of 10~mT. e) Comparison of FFT spectra of FeSn in ZF and wTF 10~mT at 80 K. f) FFT spectra of FeSn in wTF at selected temperatures.}
\label{ZF_wTFspec}
\end{figure*}

In the parent compound FeSn, we started with ZF measurements in both the SR and the NSR modes. As shown in both the time and FFT spectra of Fig.~\ref{10K_ZFSRNSR}, the oscillation amplitude observed in the NSR mode is much smaller than that in the SR mode in the antiferromagnetically ordered state. This feature indicates that the local field at the muon stopping site is pointing nearly perpendicular to the c-axis within the ab-plane. If the local field is fully perpendicular to the c-axis, we expect the oscillation amplitude in ZF-NSR to be half that in ZF-SR. As shown later, this field orientation is also consistent with the planar magnetic structure for the muon site estimated by our calculation. Figs.~\ref{ZF_wTFspec}(a) and (b) show examples of the time spectra observed in the ZF-SR mode at several selected temperature points, in a longer time range up to t = 8 $\mu s$ in (a) and a shorter time range of 0.3 $\mu s$ in (b). Fourier transforms (FFT) of the ZF-SR data shown in Fig.~\ref{ZF_wTFspec}(c) indicate that there is only one frequency in the ordered state, corresponding to a local field strength of $\sim$~0.2~T at the muon stopping site. The precession frequency develops around 250 K, well below T$_N$~$\sim$ 376~K, which is likely due to muon diffusion. A significant broadening of the FFT signal is seen below 50 K, corresponding to the fast depolarization of the precession signal in Fig.~\ref{ZF_wTFspec}(b) at low temperatures.

To check the magnetic volume fraction and any possible phase separations, we then turned to the TF configuration and applied a small transverse field 10~mT perpendicular to the c-axis and parallel to the ab-plane. Fig.~\ref{ZF_wTFspec}(d) shows examples of the time spectra observed in the TF-SR mode. Fig.~\ref{ZF_wTFspec}(e) compares the FFT spectra observed in ZF and the weak transverse field (wTF) of 10~mT at T = 80~K, and Fig.~\ref{ZF_wTFspec}(f) shows the FFT spectra obtained at several different temperatures in the wTF configuration. In the FFTs of the wTF data, two high-frequency peaks were observed at an equal distance from the ZF frequency peak, in addition to the low-frequency peak at the applied field of 10~mT. The existence of the muon precession component at the applied field frequency in wTF readily indicates that there is a muon population that resides at a site where the static internal field from the ordered moment cancels or in phase-separated regions which remain paramagnetic. We note that our powder X-ray diffraction measurements at 300~K on these crystals reveal no detectable impurity phases or indications of phase separation within the resolution of our experiments. Meanwhile in the case of Fe$_{0.89}$Co$_{0.11}$Sn, as shown in the subsequent section of the present paper, this "field absent site" disappears, presumably due to the disorder of the Fe spin network introduced by Co substitutions. This observation strongly suggests that the "field absent site" in FeSn originates from a cancellation of static local fields from the ordered Fe moments at certain muon sites.

\subsubsection{Analyses of ZF/wTF $\mu$SR: anomalies at low temperatures} \label{Analyses}

We fit the ZF time spectra with a function assuming one oscillation frequency $\gamma_{\mu} B_{int}$
\begin{equation}
\begin{split}
A(t) = A_{tot}[f_{osc} \cos (\gamma_{\mu} B_{int}t + \phi_{ZF}) e ^{-\lambda_{T} t} \\
+ (1-f_{osc}) e ^{-\lambda_{L} t}]
\end{split}
\label{ZFeq}
\end{equation}
and wTF time spectra with a function with three oscillation frequencies
\begin{equation}
\begin{split}
A(t) = A_{tot}[f_{1} \cos (\gamma_{\mu} B_{1}t + \phi_{1})e ^{-\lambda_{1T} t}\\
+f_{2}\cos (\gamma_{\mu} B_{2}t + \phi_{2})e ^{-\lambda_{2T} t}\\
+(1-f_{1}-f_{2}) \cos (\gamma_{\mu} B_{ext}t + \phi_{ext} ) e ^{-\lambda_{ext} t}]
\end{split}
\label{wTF}
\end{equation}

where $f$ and $B$ represent the volume fraction and the field strength of the oscillation components; $f_{osc}, B_{int}$ for ZF and $f_{1,2}, B_{1,2}$ for wTF. $\lambda_{T}$ and $\lambda_{ext}$ represent the transverse damping rate caused by the finite field distribution at the muon site and $\lambda_{L}$ the longitudinal relaxation rate which is also usually denoted as $1/T_{1}$. The oscillating component $f_{osc}$ in ZF comes from the spontaneous magnetic internal field $B_{int}$, while the corresponding magnetic components observed in wTF are $f_{1}$ and $f_{2}$. As shown later, $f_{osc}$ is very close to $f_{1} + f_{2}$ $\sim$~0.6. The remaining non-oscillating component $(1-f_{osc})$ in ZF in Eq.~\ref{ZFeq} can originate from a paramagnetic/non-magnetic environment for muon and/or the longitudinal asymmetry component in the magnetically ordered regions known as the "1/3 tail" for the case of a powder sample in ZF. Since the present specimens are single-crystalline with ab-plane randomly aligned, we expect a negligible '1/3 tail' component in the ZF-SR mode. In this situation, this $(1-f_{osc})$ $\sim$ $(1-f_{1}-f_{2})$ component should oscillate in wTF at a frequency of the applied external field $B_{ext}$, and this was confirmed in our experiments. The volume fraction $(1-f_{osc})$ can be due to a phase-separated paramagnetic region and/or a fraction of muons located at a high-symmetry site where the local field cancels. This aspect will be discussed later with the simulation of the muon sites and local fields.

\begin{figure*}
\includegraphics[width=0.9\textwidth]{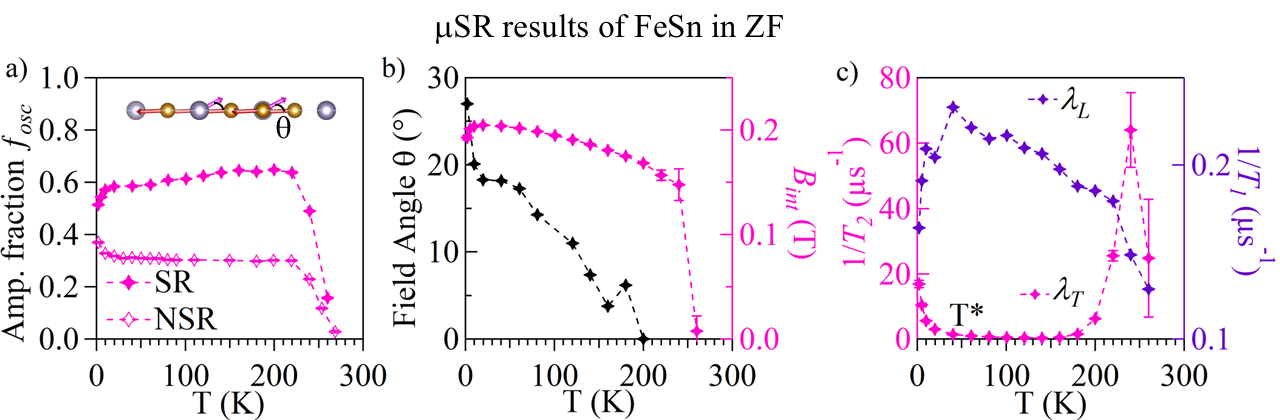}
\caption[width=\textwidth]{Fitting results extracted from Eq.~\ref{ZFeq}. a) T dependence of the oscillating amplitude fraction $f_{osc}$ in both SR and NSR mode. The inset shows the angle $\theta$ between the muon local field $B_{int}$ and the ab-plane. b) T dependence of the estimated field angle $\theta$ and the spontaneous magnetic internal field $B_{int}$. c) T dependence of the transverse relaxation rates $1/T_{2}$ and the longitudinal relaxation rate $1/T_{1}$. }
\label{ZF_fit}
\end{figure*}

In Fig.~\ref{ZF_fit}(a), we show the temperature dependence of the fraction of the oscillating signal $f_{osc}$ extracted from fitting ZF time spectra in the SR mode with Eq.~\ref{ZFeq}. This figure indicates that $\sim$ 40 $\%$ muons remain in a para/non-magnetic environment below 240 K. We also show the fraction of the oscillating signal in ZF-NSR mode, which is roughly half of the fraction seen in ZF-SR. This further confirms that the local field at the high-field muon site is pointing nearly perpendicular to the c-axis within the ab-plane. However, in a close inspection of the amplitudes in the ZF-SR and ZF-NSR modes in Fig.~\ref{ZF_fit}(a), we find that this 2:1 ratio changes below T~$\sim$~150 K. Such change can be expected if the local field at the high-field site exhibits canting toward the c-axis direction, as the oscillating amplitude follows $f_{osc}(\theta)~\sim~\cos^2{\theta}$ with the canting angle $\theta$ illustrated in the inset of Fig.~\ref{ZF_fit}(a). The canting angle estimated from the amplitude ratio is shown in Fig.~\ref{ZF_fit}(b). In Fig.~\ref{ZF_fit}(b), we also show the temperature dependence of the spontaneous magnetic internal field $B_{int}$ at the high-field muon site. The T-dependence of the frequency exhibits a departure at low temperatures from the standard behaviour expected for a sublattice magnetization of an antiferromagnet.

Fig.~\ref{ZF_fit}(c) shows the transverse relaxation rate $1/T_{2}$ of the precessing component and the longitudinal relaxation rate $1/T_{1}$ observed in the ZF-SR mode. The increase in $1/T_{2}$ below T $\sim$ 50~K (denoted as $T^{*}$) suggests some anomaly, which will be discussed later with $\mu$SR results in LF and the corresponding anomaly found in magnetic susceptibility. In contrast, $1/T_{2}$ peak at T $\sim$ 250~K is not associated with any corresponding signature of susceptibility and therefore can be attributed to the effect of muon diffusion.

\subsubsection{Double-peak spectra in wTF}
\begin{figure*}[h]
\includegraphics[width=\textwidth]{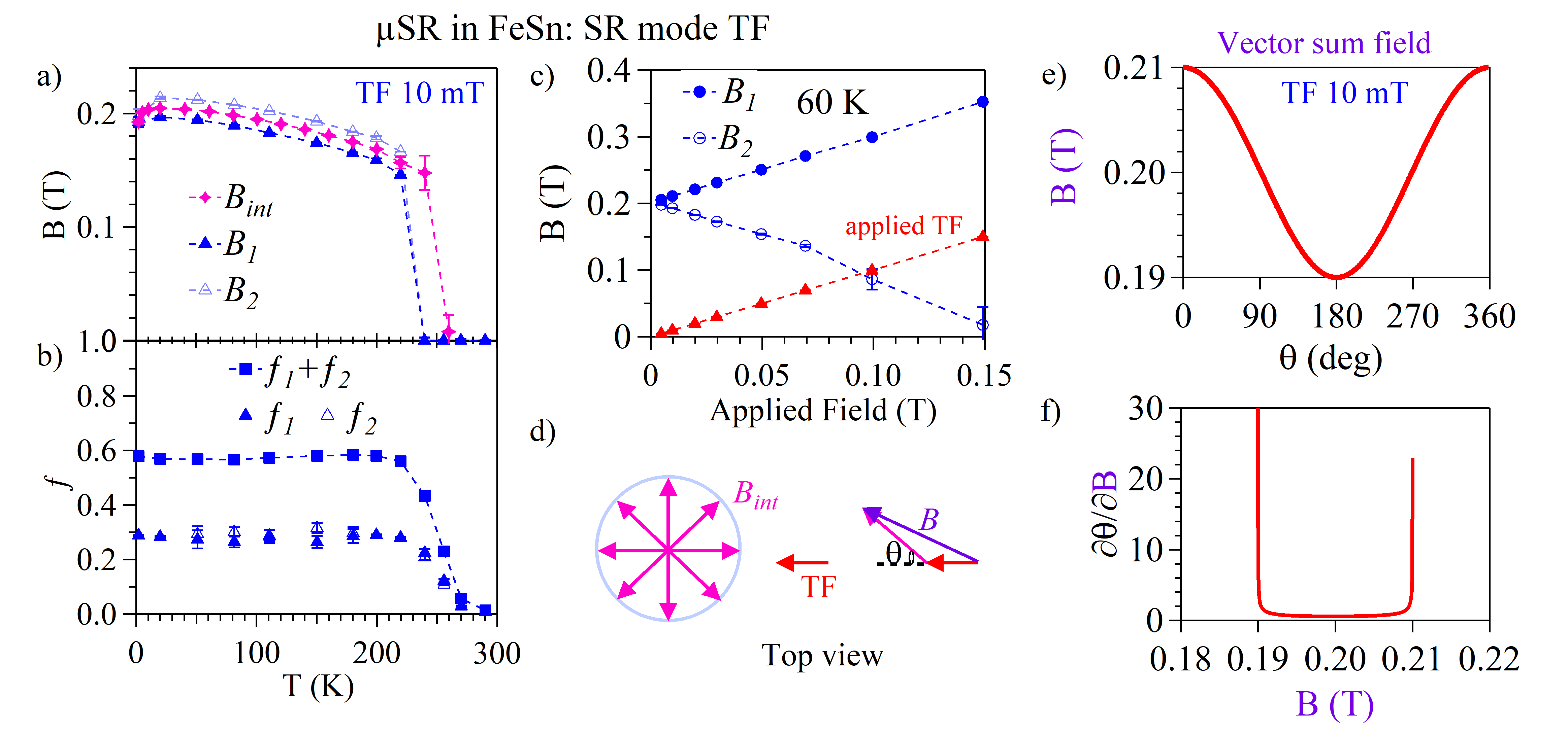}
\caption[width=\textwidth]{a) T dependence of the magnetic fields, b) T dependence of volume fraction under TF 100G. c) Transverse field dependence of the observed frequencies, $B_{1,2}$ of FeSn at 60 K. d) Top view of the Fe magnetic moment in the ab-plane, $B_{int}$, external field TF was applied within the plane, and the vector sum, $B$, of the internal and external fields. e) The simulated angular dependence of the field strength f) The probability of the field seen by muons, as described in the main text.}
\label{SR_TF}
\end{figure*}

In our wTF measurements, we notice that the high-field muon site with $B_{int}$ in ZF generates oscillating signals with two split frequencies close to $B_{int}$, as shown in Figs.~\ref{ZF_wTFspec}(e) and (f). We denote these fields as $B_{1}$ and $B_{2}$, and their signal amplitude fraction as $f_{1}$ and $f_{2}$, and show the temperature dependencies in Figs.~\ref{SR_TF}(a) and (b). In wTF = 10~mT, $B_{1}$ and $B_{2}$ are very close to the ZF spontaneous magnetic field $B_{int}$ plus or minus 10~mT and their fractions are equal to each other $\sim$ 0.3 independent of temperature below T $\sim$ 250 K. In Fig.~\ref{SR_TF}(c), we show how these two frequencies depend on the varying transverse external field $B_{ext}$ at T = 60 K. This figure demonstrates the linear relationships between $B_{1}$ and $B_{2}$ on $B_{ext}$. In the wTF geometry, the single crystals are aligned by their c-axis, while directions of a-or-b axes are randomly oriented within the ab-plan. The external TF field was applied perpendicular to the c-axis. The local field observed in this configuration should be a vector sum of the internal field and the external field, $\vec{B_{int}} + \vec{B_{ext}}$, as illustrated in Fig.~\ref{SR_TF}(d) which shows a top view of the vector sum. The angular dependence of the local field strength $\vec{B}$ seen by the muons can be derived from ${B^{2} = B^{2}_{int} + B^{2}_{ext} + 2B_{int}B_{ext}\cos(\alpha)}$, where $\alpha$ is the angle between $B_{int}$ and $B_{ext}$. By assuming $B_{int}$ = 0.2~T, we plot the angular dependence of the field $B$, in Fig.~\ref{SR_TF}(e). The probability of fields seen by muons, $\frac{\partial \alpha}{\partial B}$ in Fig.~\ref{SR_TF}(f), can then be used to describe the field distribution in the time spectra. This is consistent with the two peaks of $B_{1}$ and $B_{2}$ observed in the FFT of the wTF spectra in Figs.~\ref{ZF_wTFspec}(e) and (f).

\subsubsection{Low temperature anomaly and dynamic responses in LF}\label{LF-muSR}

\begin{figure}
\centering\includegraphics[width=0.75\columnwidth]{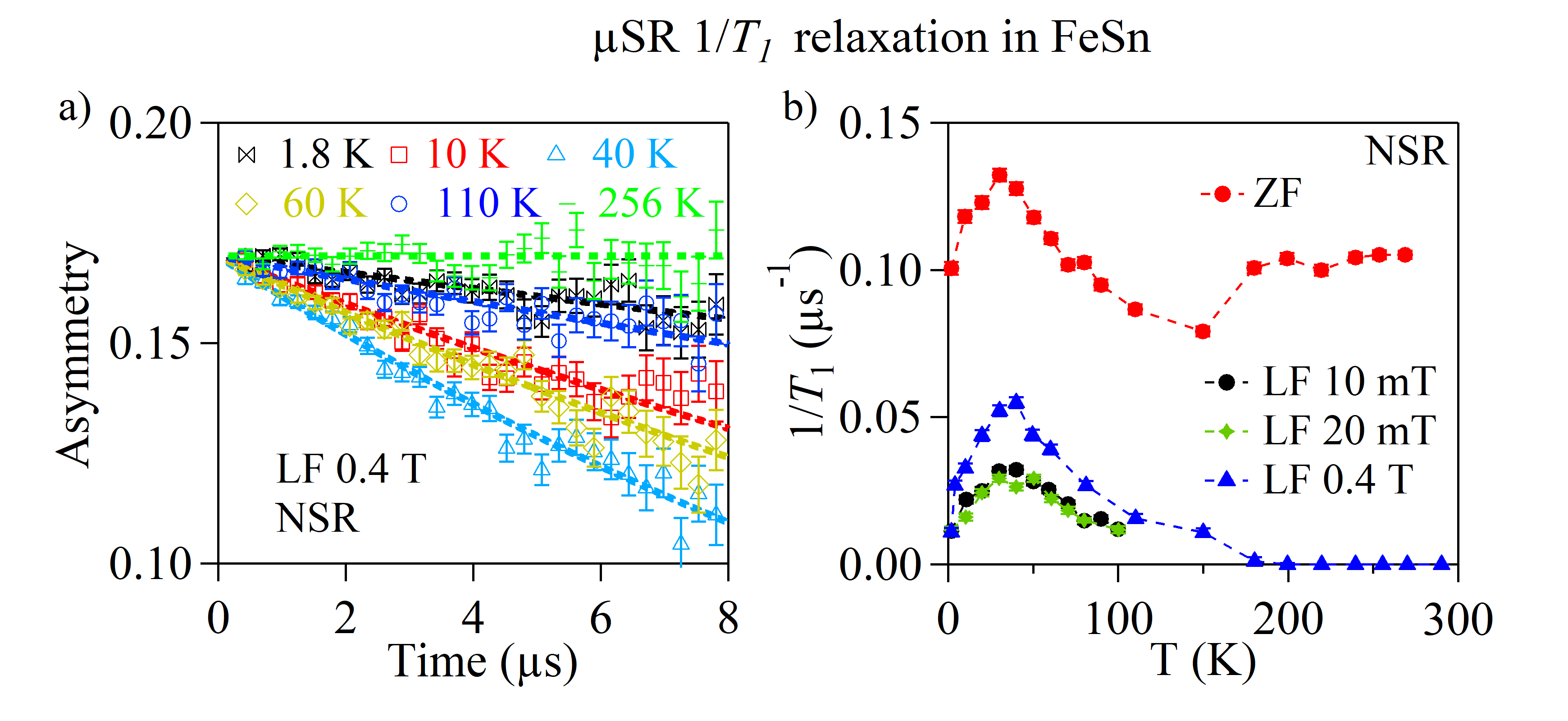}
\caption{a) LF-0.2~T time spectra at selected temperature in NSR mode, b) Temperature dependence of the longitudinal relaxation rate, $1/T_{1}$, in ZF, LF-10~mT, LF-20~mT and LF-0.4~T. A relaxation peak in $1/T_{1}$ is robustly seen below 50 K.}
\label{T1}
\end{figure}

To further study the anomalies found in ZF at low temperatures, as described in \ref{Analyses}, we performed ZF and LF $\mu$SR measurements in the NSR mode, where external fields were applied parallel to the kagome plane and the initial muon spin. Fig.~\ref{T1}(a) shows time spectra at selected temperatures under LF-0.4~T. In low LF with 10~mT and 20~mT, we fit the time spectra with Eq.~\ref{ZFeq} as used in ZF. We fit the LF-0.4~T spectra with a simple exponential relaxing function: $A(t)= A_{tot}e ^{-\lambda_{L,LF4kG} t}$ in the time range of 0.5-10 $\mu$s. In Fig.~\ref{T1}(b), we show the relaxation rate observed in ZF and LF under NSR configuration. A clear $1/T_{1}$ peak was observed at T~$\sim$ 50~K both in ZF and LF measurements, suggesting that dynamic spin fluctuations taking place well below the antiferromagnetic ordering temperature 376~K of FeSn. The change of the time spectra in Fig.~\ref{T1}(a) involves at least half of the full amplitude, which indicates that this is a change taking place in the bulk volume fraction, ruling out the possibility of an effect originating from a minority impurity phase. Together with the reorientation and broadening of the static internal field discussed in Sec.~\ref{Analyses}, these anomalies detected by $\mu$SR at low temperatures suggest the existence of a magnetic instability in bulk FeSn.

\subsection{Magnetic susceptibility and resistivity of FeSn} \label{STMsec}

\begin{figure}[H]
\includegraphics[width=\columnwidth]{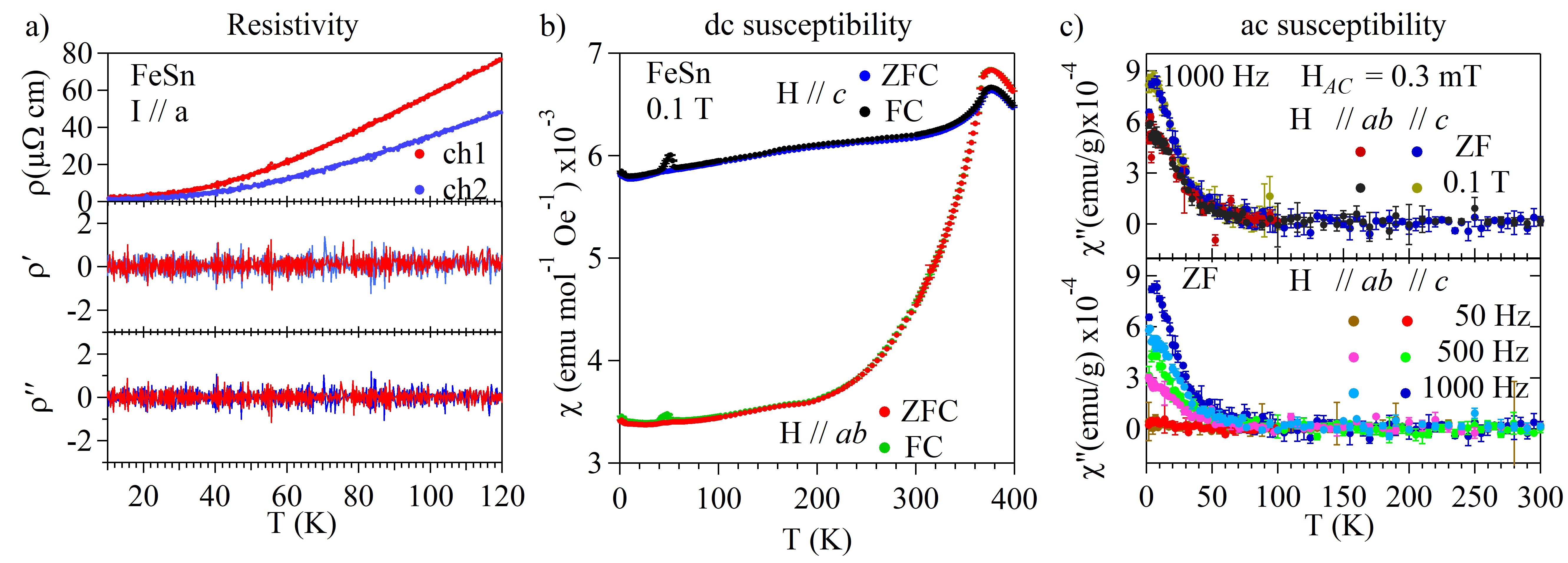}
\caption{a) T dependence of resistivity and its derivatives in two independent channels setup ch1 and ch2 on the same crystal. Current, $I$, was applied along a direction. b) dc susceptibility results of FeSn under 1 kG external field along the c-axis and ab-plane. In addition to the antiferromagnetic transition at $\sim$ 376 K, a small kink around $\sim$~50 K was observed in the ZFC mode which was further enhanced in the FC mode. c) Frequency and field dependence of ac susceptibility in FeSn with ac amplitude of 3 G. Additional dc field was applied both along the c-axis and ab-plane. The observed imaginary part $\chi''$ is independent with the field while exhibiting a clear increase below 50 K.}
\label{res_sus}
\end{figure}

In the previous transport studies on FeSn, no clear sign of any anomaly below 50 K was reported \cite{Sales2019PRM, Meier2020_PRB, Liu2020, Kang2020, Kang2020_2, Meier2019, Kakihana_JPSJ2019, Sales2021}. To search for a possible signature corresponding to the low-temperature anomaly found in our $\mu$SR results, we performed resistivity measurements. As shown in Fig.~\ref{res_sus}(a), no clear signature of any anomaly was found in resistivity and its temperature derivatives.

We also performed dc-and-ac magnetic susceptibility measurements. Fig.~\ref{res_sus}(b) shows the dc susceptibility measured under external field H of 0.1~T. The results with H along the c-axis correspond to the perpendicular susceptibility, while H perpendicular to the c-axis includes an average 50~$\%$ of parallel and 50~$\%$ of perpendicular susceptibility in the absence of spin flopping. The observed amplitude ratio of these two configurations is consistent with the expectation of an antiferromagnet with a transition temperature of $\sim$ 376 K.
In addition to this antiferromagnetic transition, a small kink at T~$\sim$~50~K was also observed in the ZFC mode which was further pronounced in the FC mode. Upon closely examining the previously reported temperature dependence of the derivative of dc susceptibility in Ref.~\cite{Sales2019PRM}, we find a similar kink corresponding to this anomaly.

Fig.~\ref{res_sus}(c) shows the results of our ac susceptibility measurements at several frequencies with the driving ac field both parallel and perpendicular to the c-axis. A clear increase in the imaginary part $\chi''$ was observed below 50~K, for both driving field direction in zero external field and the 0.1~T external field applied along the c-axis and the ab-plane. This result indicates a development of magnetic dynamics at low temperatures below T~$\sim$~50~K. These susceptibility results correspond very well to the anomalies observed in $1/T_{1}$ and $1/T_{2}$ by $\mu$SR. Note that $1/T_{1}$ and $\chi''$ both reflect dynamic spin fluctuations. This excellent correspondence indicates that the $\mu$SR anomalies observed at and below T~$\sim$~50~K are most likely due to the intrinsic property of FeSn, rather than a phenomenon caused specifically by the existence of a positive muon in the material.

\subsection{Scanning tunneling microscopy and spectroscopy of FeSn}

\begin{figure}[H]
\includegraphics[width=\columnwidth]{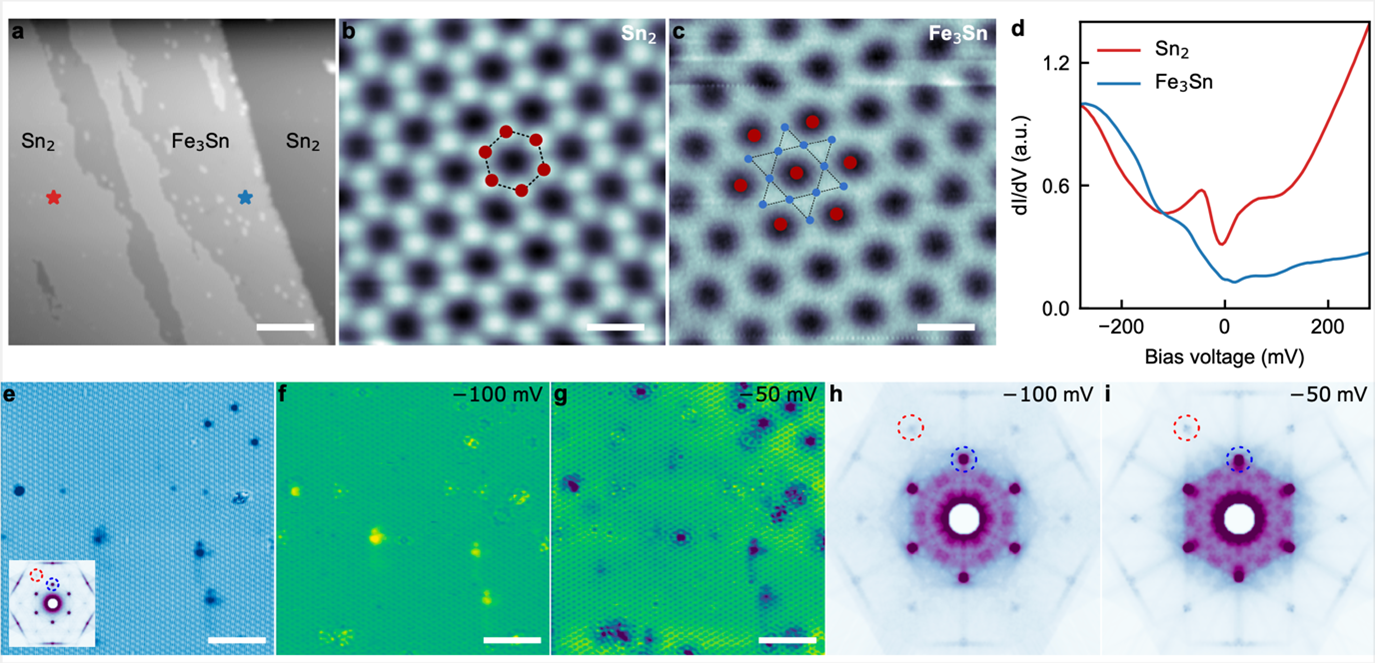}
\caption{STM on the bulk cleaved surface of FeSn. a) STM topography of the cleaved surface showing Sn$_2$ and Fe$_3$Sn termination layers. b) Constant current mode topography of the Sn$_2$ surface showing a honeycomb structure where individual Sn atoms can be seen. c) Constant current topography of the Fe$_3$Sn kagome termination with a lattice model overlaid on top. Red atoms are Sn and blue atoms are Fe. d) Scanning tunnelling spectroscopy on the surfaces of Sn$_2$ and Fe$_3$Sn, indicated in a) and e). Large area constant current mode topography of the Sn$_2$ surface showing different kinds of defects. The inset shows the FFT of the topography. f), g) Spatial map of differential conductance, dI/dV at energies -100 mV and -50 mV. h), i) The corresponding FFTs of the dI/dV map as shown in f) and g). STM setpoint conditions: a (-600 mV, 100 pA), b (50 mV, 1.9 nA), c (-200 mV, 1.9 nA), d (-300 mV, 500 pA), e (-100 mV, 200 pA), f (-100 mV, 200 pA), g (-50 mV, 200 pA). Scale bars: a (10 nm), b,c (0.5 nm), e, f, g (5 nm).}
\label{STM}
\end{figure}

We also performed scanning tunnelling microscopy and spectroscopy (STM/STS) studies on the bulk FeSn. Our main motivation comes from a previous report of STM studies on a thin film of FeSn, which included a signature of spontaneous 'trimer formation' of the kagome lattice at low temperatures~\cite{Zhang2023}, reminiscent of spin Peierls transition in low-dimensional systems. Trimer formation might explain the T $\sim$ 50 K signatures observed by $\mu$SR and magnetic susceptibility.

Figure~\ref{STM}(a) shows a large area topography of the cleaved FeSn surface. We see two distinct surface terminations, Sn$_2$ and Fe$_3$Sn, the former being the predominant surface termination. Figures~\ref{STM}(b) and (c) show atomically resolved STM topographies of the two surfaces. For the Sn$_2$ surface, we can resolve each Sn atom which shows a clear honeycomb lattice structure. For the Fe$_3$Sn kagome structure, however, the individual atoms are not resolved. A lattice model is overlaid on both topographies. Scanning tunnelling spectroscopy shows a peak at around -50 mV for the Sn$_2$ termination which is not present for the Fe$_3$Sn kagome termination \cite{Li2022, Multer2023}. To investigate the possibility of any broken symmetry visible to STM at low temperatures, we performed spectroscopic imaging experiments on both surfaces. Shown in Fig.~\ref{STM}(e) is a large area topography of the Sn$_2$ surface that has several defects. Simultaneously obtained dI/dV maps at two different energies are shown in Fig.~\ref{STM}(f) (-100 mV) and Fig.~\ref{STM}(g) (-50 mV). The corresponding FFT magnitudes are shown in Figs.~\ref{STM}(h) and (i). Our spectroscopic imaging experiments reveal no clear evidence for additional Bragg spots that have been seen in FeSn thin films \cite{Zhang2023}, and no evidence for rotational symmetry breaking.

\begin{figure}
\includegraphics[width=\columnwidth]{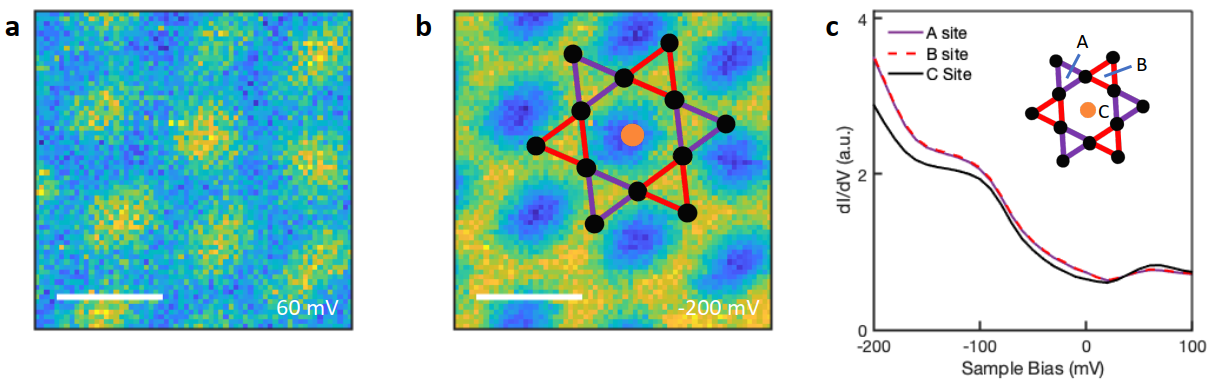}
\caption{STS results of the dI/dV from the Fe$_3$Sn termination of a cleaved bulk single crystal of FeSn. (a) dI/dV map with the bias voltage of 60 meV. (b) with -200 meV. and (c) bias voltage dependence of the dI/dV signal from A, B, and C locations as illustrated in (b). Set point conditions for a, b, and c: 300 mV, 700 pA. In contrast to a previous report on STS of a thin film specimen of FeSn~\cite{Zhang2023}, we found no difference between the signal from A and B, which rules out the formation of a trimer state.}
\label{dI-dV}
\end{figure}

The previous STM study on the FeSn film, which reported a signature of 'trimer formation', presented dI/dV signal from two adjacent triangles in Fe$_3$Sn termination. We generated a similar plot and show the results in Fig.~\ref{dI-dV} with the dI/dV map for the sample bias voltages of 60 mV in (a) and -200 mV in (b), and the averaged dI/dV signal in (c) from several locations of A, B and C, as illustrated in (b). Unlike the previous report on a thin film FeSn sample that exhibited clear differences in the signals between A and B, our results from a bulk signal crystal of FeSn showed no difference between A and B, indicating that there is no signature of 'trimer formation' in our STM results. Thus, our initial attempt to find signatures of possible phase transition results in a negative answer.
However, since STM gives selective information on the surface, this result does not necessarily represent the situation in bulk FeSn. Therefore, trimer formation has not yet been completely ruled out as a possible explanation of the T $\sim$ 50 K anomaly observed by the two bulk sensitive probes, $\mu$SR and magnetic susceptibility. We will continue STM studies at higher temperatures and start high-resolution X-ray studies to further seek the possible origin of the 50 K anomaly.

\subsection{Muon site and local fields in FeSn}\label{sec_muonsite}
To better understand the origin of the 40 $\%$ paramagnetic-like volume in FeSn, we turned to investigate the possible muon sites. The $\mu$SR results in ZF and wTF described above indicate that the static local field at the magnetic muon stopping site is pointing perpendicular to the c-axis in the antiferromagnetic state, which provides a strong restriction on the muon stopping site in the planar spin structure. Generally, the main characteristics of the coupling between the dipolar moment of Fe and the muon are the isotropic hyperfine coupling and the dipolar field coupling. 

\begin{figure}
\centering\includegraphics[width=0.4\columnwidth]{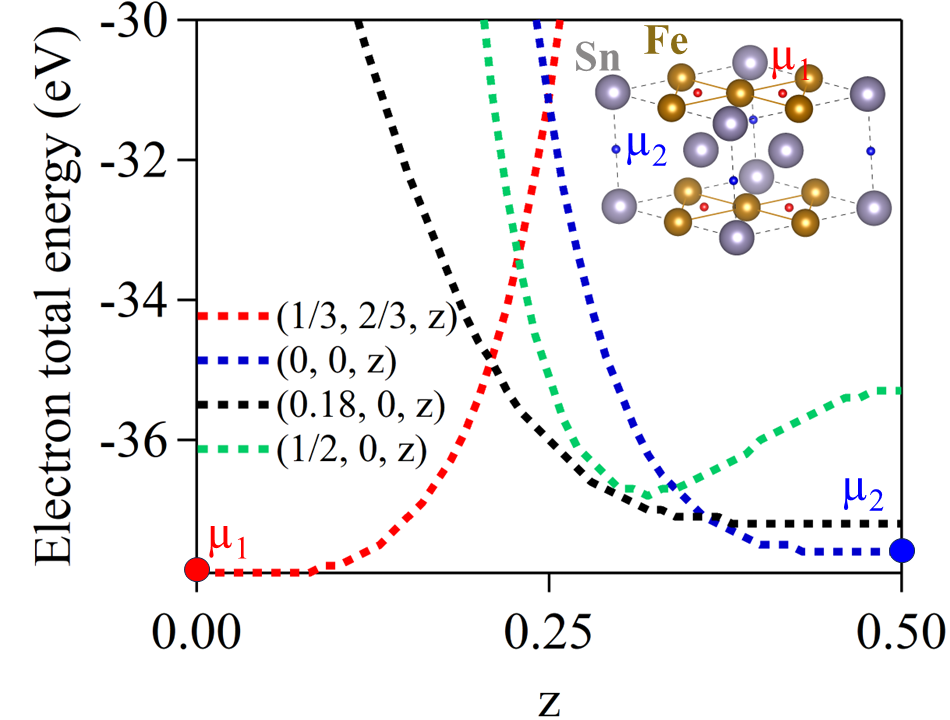}
\caption{The electron total energy of calculated muon stopping sites. Starting from a high symmetry point and moving along the c direction. Inset shows the two plausible muon stopping sites $\mu_1$ and $\mu_2$, marked as red and blue dots. The $\mu_1$ site sits in the centre of three Fe atoms while the $\mu_2$ site sits in the middle of two Sn atoms between kagome layers.}
\label{muonsite}
\end{figure}

In Fig.~\ref{muonsite}, we show the total electron energy of possible muon stopping sites, starting from a high symmetry point and then moving along the c-axis. This calculation and stability / symmetry analysis reveal that two plausible muon stopping sites are the lowest electrical potential site, $\mu_1$ = (1/3, 2/3, 0) and the second lowest site, $\mu_2$ = (0, 0, 1/2) as marked by red and blue dots symbols. The latter $\mu_2$ site is located at the high symmetry point, where the local field vanishes in the planar spin structure. This site is the strongest candidate for the field-cancelling site where $40\%$ of the implanted muons reside. The former $\mu_1$ site exists in the kagome plane, and our dipolar field calculation indicates a local field strength of 0.9352~T within the ab-plane for the planar magnetic structure with an Fe moment size of 1.85~$\mu_{B}$ as determined in Ref.~\cite{Sales2019PRM} at 100~K. This dipolar field strength is much larger than what we observed $\sim$ 0.2~T. Note that FeSn is metallic and the muon hyperfine contact field is not negligible. Previous DFT calculations and muon studies in pure bcc Fe crystal indicate a large hyperfine field $\sim$ -1.11~T~\cite{NISHIDA1977235, Onuorah2018}. The $\mu$-Fe distance in bcc Fe is about 1.587~$\AA$, which is close to the value for $\mu_1$ (1.53~$\AA$) in FeSn. By including this hyperfine field in our dipolar calculation, the estimated local field becomes 0.175~T, which agrees fairly well with our ZF-$\mu$SR observation of 0.195~T at 100~K. We can alternatively estimate the hyperfine field from the observed results in FeSn to be $\sim$ -1.07~T. 

We further studied the effect of the reorientation of Fe moments to understand the observed field canted towards the c-axis as shown in Fig.~\ref{ZF_fit}(b). Assuming the Fe spins reorient out-of-plane while maintaining the antiferromagnetic arrangement between the kagome layers, we found that a 2-degree rotation of Fe spins toward the c-axis would result in a field canted angle $\theta$ $\sim$ 32 degrees. 

\subsection{$\mu$SR in Fe$_{0.89}$Co$_{0.11}$Sn}\label{sec_Fe89}
\begin{figure}
\centering\includegraphics[width=0.7\columnwidth]{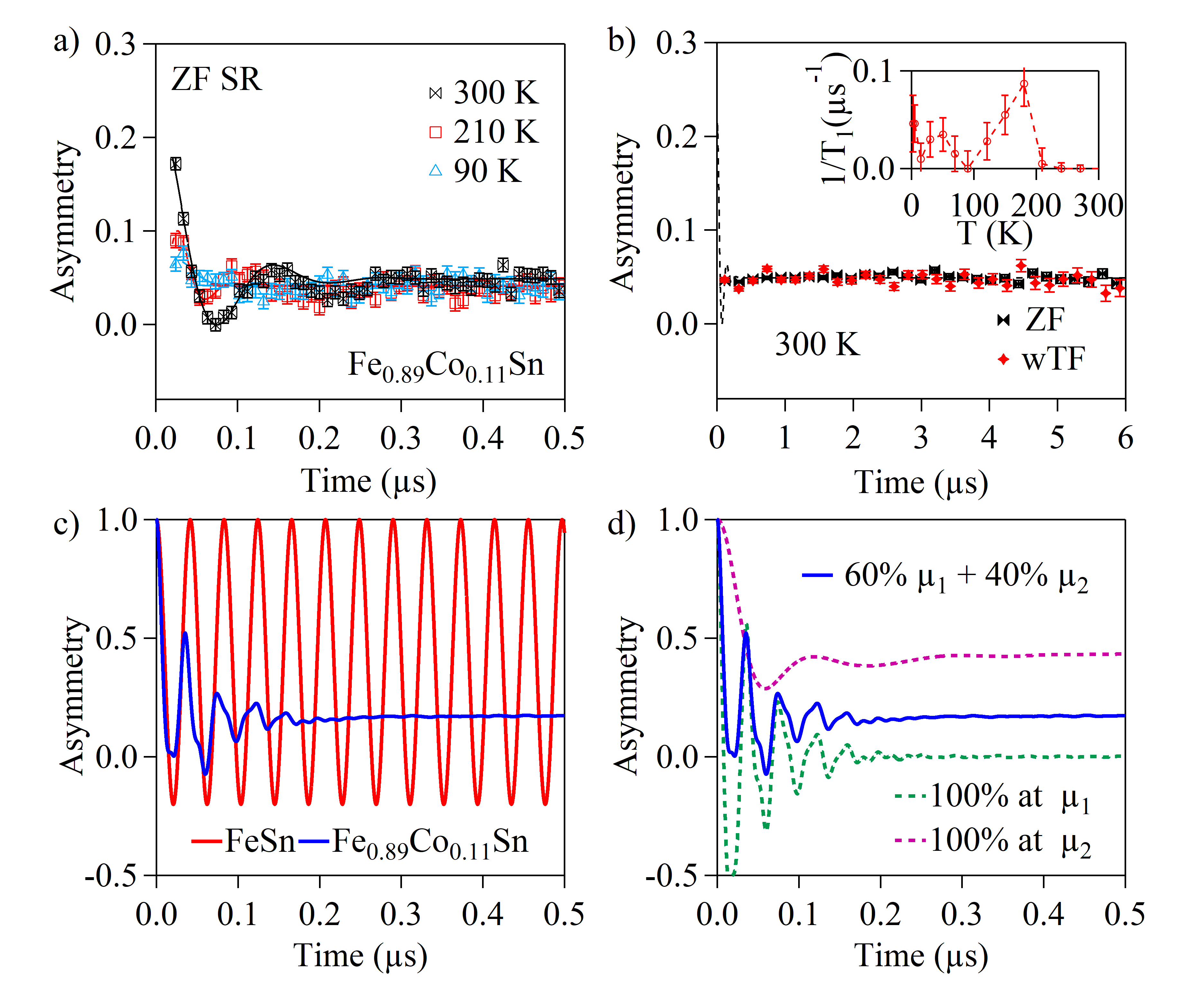}
\caption{a) ZF $\mu$SR spectra within 0.5 $\mu$s at selected temperatures. The oscillation seen at 300 K was smeared out by the fast $1/T_{2}$ rate with a decrease in temperature. b) wTF 10~mT time spectra in the long time window, which shows no oscillation, indicating a fully ordered magnetic state at 300 K. The inset shows the temperature-dependent small $1/T_{1}$ relaxation rate, which exhibits a transition at around 200 K. c) Simulation of ZF time spectra for FeSn and Fe$_{0.89}$Co$_{0.11}$Sn. d) ZF time spectra of Fe$_{0.89}$Co$_{0.11}$Sn with different mixture volume of suggested muon sites.}
\label{Fe89}
\end{figure}

To further examine the origin of the 40~$\%$ volume fraction without static internal field (except for nuclear dipolar fields) observed in FeSn, we turned to Co-doped Fe$_{0.89}$Co$_{0.11}$Sn. In Figs.~\ref{Fe89}(a) and (b), we show the time spectra of Fe$_{0.89}$Co$_{0.11}$Sn in ZF-SR at selected temperatures and TF 10~mT at 300~K. A fast-damped oscillation at 300~K in ZF was observed, indicating behaviours expected for a full volume of static magnetic order. With decreasing temperatures, oscillations seen at 300~K turn into a fast damping signal, which may be due to the reorientation of Fe moments from a planar magnetic structure to a tilted complex magnetic structure~\cite{Meier2019, Sales2021}. We present the temperature dependence of the longitudinal relaxation rate in the inset of Fig.~\ref{Fe89}(b). A peak in the $1/T_{1}$ rate was observed at about 200~K, corresponding to the magnetic transition as seen in the phase diagram in Fig.~\ref{Phasediagram}.

The time spectra at TF 10~mT in Fe$_{0.89}$Co$_{0.11}$Sn in Fig.~\ref{Fe89}(b) do not show any oscillations at the applied TF. This implies that there is no paramagnetic or nonmagnetic volume, and the full 100~$\%$ volume is in a magnetically ordered state at 300~K. This absence of phase separation between magnetically ordered and paramagnetic volumes in the Co 11~$\%$ compound strongly suggests that the non-oscillating signal with 40~$\%$ volume fraction observed in FeSn is not due to phase separation, although the situation in the Co 11~$\%$ compound does not directly represent that in pure FeSn.

We simulated the ZF time spectra of Fe$_{0.89}$Co$_{0.11}$Sn based on the dipolar and hyperfine fields by randomly replacing 11~$\%$ Fe atoms with non-magnetic Co atoms while retaining the planar magnetic structure of FeSn. Although the hyperfine field may vary with different concentrations and materials, we used the hyperfine field of $\sim$ -1.11~T in our simulation as the simplest approximation. Figure~\ref{Fe89}(c) shows the ZF time spectra of Fe$_{0.89}$Co$_{0.11}$Sn and FeSn, generated by assuming a volume ratio of $ 60~\%~: 40~\%$ for $\mu_1$ and $\mu_2$ sites. The depolarization in Fe$_{0.89}$Co$_{0.11}$Sn increases as compared to that in FeSn, which is consistent with our experimental results.

We further generated hypothetical ZF time spectra for Fe$_{0.89}$Co$_{0.11}$Sn, as shown in Fig.~\ref{Fe89}(d), assuming muon stopping sites to be exclusively either $\mu_1$ or $\mu_2$ site. The simulated spectrum with 100$\%$ $\mu_2$ site (purple broken line) shows fast depolarization followed by a 1/3 tail. The field-absent phenomenon disappears because the local high symmetry was broken due to the disorder caused by the (Fe, Co) substitutions. This simulation provides further support to our argument that the 40 $\%$ field-absent volume fraction in FeSn is due to internal field cancellation at the high-symmetry muon site $\mu_2$, rather than the effect of phase separation.

\subsection{$\mu$SR in Fe$_{0.2}$Co$_{0.8}$Sn}{\label{Fe20section}}
\subsubsection{ZF and LF $\mu$SR results}

\begin{figure}
\includegraphics[width=\columnwidth]{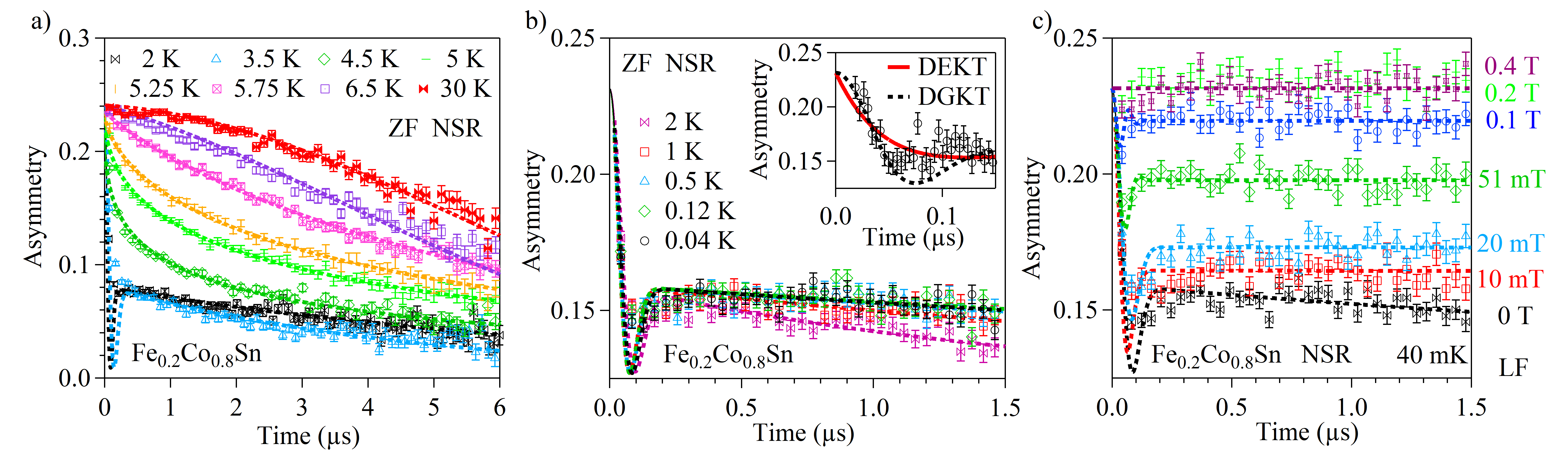}
\caption{Dashed lines are fitting results as described in the main test. a) ZF-$\mu$SR time spectra of Fe$_{0.2}$Co$_{0.8}$Sn at selected temperature above 2~K. b) ZF-$\mu$SR time spectra with temperatures below 2~K. Inset: the front end of the 40~mK time spectra with the fitting of both Exponential relaxation and Gaussian relaxation. c) LF-$\mu$SR spectra at 40~mK with fields up to 0.4~T. The relaxation rate was quickly decoupled and fully decoupled above 0.2~T.}
\label{Fe20specs}
\end{figure}

As shown in the phase diagram in Fig.~\ref{Phasediagram}, Co substitutions above 50~$\%$ lead to a spin glass state. In this region, we performed $\mu$SR studies of Fe$_{0.2}$Co$_{0.8}$Sn with the spin freezing temperature $T_g \sim 3.5$~K \cite{Sales2021}. Figure~\ref{Fe20specs}(a) shows time spectra of ZF NSR measurements taken at M20 at T $\geq$ 2~K with the dashed lines representing fits to Eq.\ref{M20Spinglass}. At very high temperatures, such as T = 30~K, the effect of the local field from Fe moments is eliminated by the fast spin fluctuations. Here, muon depolarization is caused by nuclear dipolar fields (mainly from Co nuclei in the present case), and we observed a Gaussian Kubo-Toyabe function $g_{z}(\Delta, t)$ for static nuclear fields with $\Delta$(30~K) = 0.121~$\mu s^{-1}$. With a decrease in temperature, the spectra changed into a product of the dynamic relaxation because of the fluctuating local field from Fe and the above-mentioned static Kubo-Toyabe function. Upon approaching $T_g$, the dynamic depolarization rate rapidly increases, suggesting the slowing down of the Fe moments. In general, line shapes for dynamic relaxation are given by stretched exponential functions, $e^{-(\lambda t)^\beta}$, with the stretching power $\beta$ depending on the field distributions and fluctuation time scales. For example, $\beta = 1/2$ can be expected for diluted and random magnetic systems in the narrowing limit \cite{Hartmann_1968, Uemura1985}. The present system has relatively dense Fe moments, as suggested by the Gaussian initial damping of the time spectrum at T$\rightarrow$0 shown in the inset of Fig.~\ref{Fe20specs}(b). In this case, a simple exponential decay with $\beta = 1$ is expected for the dynamic relaxation function. The observed time spectra agree well with this expectation, as shown by dotted lines in Fig.~\ref{Fe20specs}.

With further decreases in temperatures below $T_{g}$, the polarization rapidly drops to~$\sim$~1/3 of its initial value followed by the 1/3 tail, which slowly relaxes in the longer times as expected for coexisting static and dynamic random local fields. We used a phenomenological model to extract the static and dynamic effects of local fields using a product function of the static response of the Kubo-Toyabe function with dynamic simple exponential decay. The observed 1/3 tail relaxes slower at 2~K as compared to 3.5~K as shown in Fig.~\ref{Fe20specs}(a), which indicates a decrease of dynamic relaxation in lower temperatures. To characterize the temperature dependence of the dynamic relaxation, we performed additional measurements on the same crystals down to 40~mK using the M15 beamline. The time spectra observed in ZF-NSR mode in Fig.~\ref{Fe20specs}(b) demonstrate that the 1/3 tail relaxes even slower at lower temperatures. The early-time relaxation of the 40~mK data, as shown in the inset, was better described by a Gaussian relaxation~(DGKT) rather than an exponential relaxation~(DEKT). This indicates that the local field distribution is close to Gaussian, as expected in relatively dense magnetic systems. 

We also performed LF $\mu$SR measurements, where the external field is applied in the direction of the initial muon spin. A static signal will be nearly fully decoupled by an applied field that is a few times larger than the static internal field measured in zero field. In contrast, if the relaxation of the ZF $\mu$SR signal is due to dynamics, the signal will not be decoupled by an applied field of this magnitude. Figure~\ref{Fe20specs}(c) shows the LF results at 40~mK. The 1/3 tail becomes nearly flat (non-relaxing) with a small LF of only 10~mT, indicating that the dynamic effect is negligible. We also observed a gradual decoupling of fast relaxation with increasing applied longitudinal fields and a nearly complete restoration of asymmetry with LF~$\sim$~0.2~T, consistent with the initial ZF relaxation shown in the inset of Fig.~\ref{Fe20specs}(b).

We shall note that the background signal from the sample holder and surrounding sample environment in the dilution cryostat gives a significant contribution (up to~$\sim$~40~$\%$ of the signal amplitude) to the data taken at M15. The slow but finite decay of the flat signal observed in ZF at 40~mK shown in Fig.~\ref{Fe20specs}(b), is due to the effect from the nuclear dipolar field in the background signal, which was eliminated by applying a small external LF of 10~mT. To fit the signal from the sample, we assumed the phenomenological function $G_{SKT}(\Delta,t)$ multiplied by dynamical simple exponential decay rate $1/T_1$ as,

\begin{equation}
 A(t) = A_{tot} G_{SKT}(\Delta,t) e^{-\frac{1}{T_1} t}
\label{M20Spinglass}
\end{equation}
where 
\begin{equation}
G_{SKT}(\Delta,t) = \frac{1}{3} + \frac{2}{3} ( 1 - (\Delta t)^ \alpha ) e^{-(\Delta t)^\alpha /\alpha}
\label{SKT}
\end{equation}
with $\alpha$ = 2 for the Gaussian Kubo-Toyabe function:
\begin{equation}
g_{z}(\Delta, t) = \frac{1}{3} + \frac{2}{3} ( 1 - (\Delta t)^2 ) e^{-(\Delta t)^2 /2}
\label{GKT}
\end{equation}

\begin{figure}
\includegraphics[width=\columnwidth]{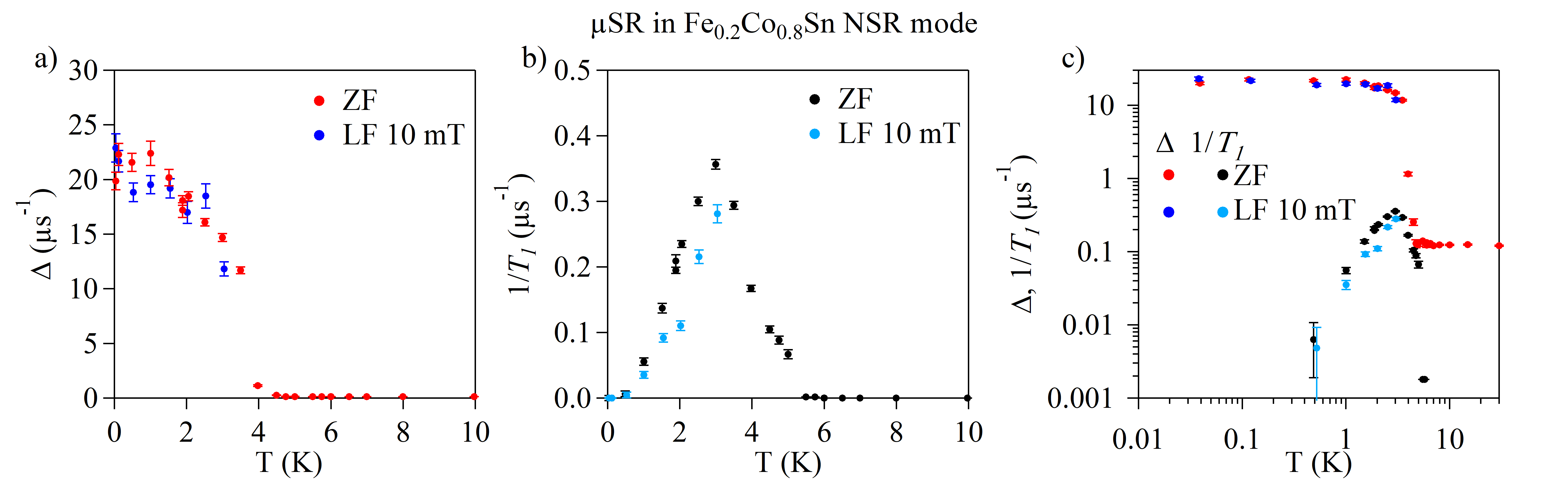}
\caption{T-dependence of the a) static relaxation rate $\Delta$ and b) dynamical fluctuation rate, $1/T_1$ at ZF/LF-100G in NSR mode. c) Plot of $\Delta$ and $1/T_1$ together on a logarithmic scale.}
\label{Fe20ZFfit}
\end{figure}

The background signal in the M15 data was included by an additional term $e ^{-(\lambda_{bg} t)}$ in the fitting process. Since the M20 apparatus is an ultra-low background setup, we compared the M15 data and M20 data taken at the same temperatures to correctly obtain the sample volume fraction and its slow fluctuation rate. Fig.~\ref{Fe20ZFfit} shows the temperature dependence of the relaxation rate, $\Delta$, and the low temperature dynamical fluctuation rate, $1/T_1$, thus derived from this fit.

In spin glass systems, the relaxation function in ZF around $T_{g}$ often exhibits complicated line shapes, which are difficult to parameterize with a simple formula. Given this, we adopted an approach to use the general Kubo-Toyabe function, connecting smoothly crossing $T_{g}$. In the region at temperatures below $T_{g}$, the stretched power parameter $\alpha$ was set to 2 to account for the observed early Gaussian decay. The static relaxation rate, $\Delta$, which reflects the width of the static internal field distribution, can be viewed as the order parameter. The Gaussian relaxation rate at 40~mK gives an estimated static local field as $\Delta /\gamma_{\mu} =20.28/(2 \pi \times 135.5) = 23.8$~mT. We note that the minimum of the Kubo-Toyabe function does not agree perfectly with our data, as shown in Fig.~\ref{Fe20specs}, which might mean that the true field distribution is somewhat more complicated. The dynamical fluctuation rate, $1/T_1$, peaks around $T_{g}$ and decreases to 0.001(4)~us$^{-1}$ at 40~mK in ZF. We took additional LF 10~mT measurements to study the dynamic relaxation rate. The fluctuation rates $1/T_1$ in LF 10~mT decrease with temperature and becomes essentially zero below our detection limit at T = 40~mK. These results indicate that the dynamic spin fluctuations are frozen and the spin system becomes completely static. These behaviours are similar to previously known results in the canonical dilute-alloy spin glasses AuFe and CuMn. \cite{Uemura1985}, while distinctly different from the 'persistent spin dynamics' observed in several geometrically frustrated spin systems at T~$\rightarrow$~0.

\subsubsection{Fe moment size estimated from comparisons between the observed spectra and the dipolar field simulation}

\begin{figure}[h]
\includegraphics[width=\columnwidth]{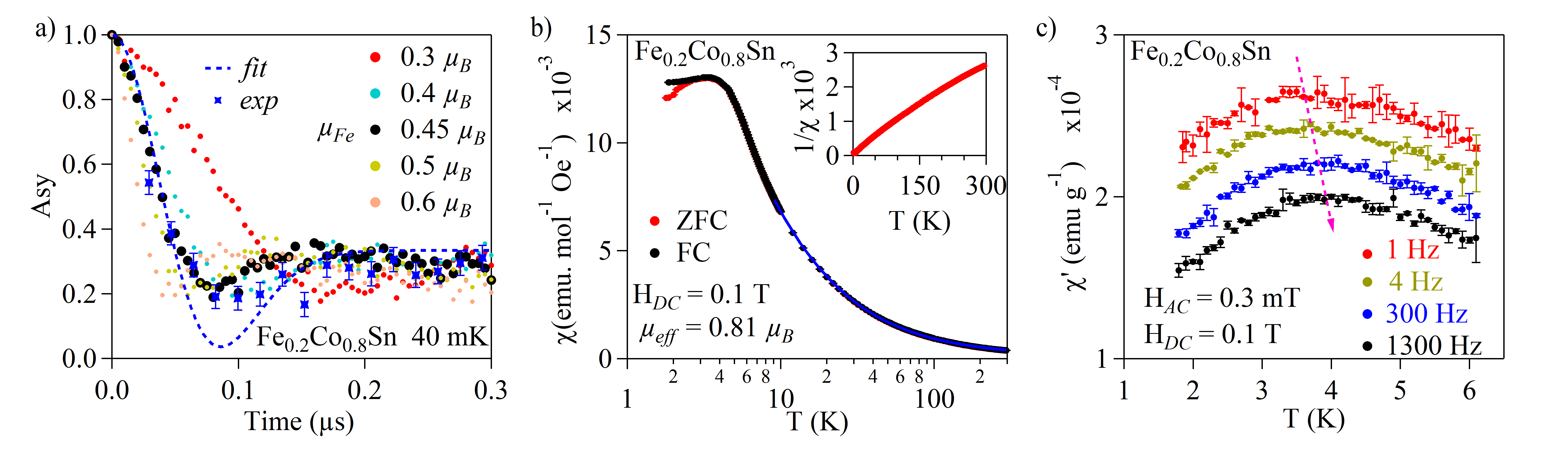}
\caption{a) ZF time spectra with various Fe moment sizes from dipolar field simulations. The blue line is the fitting result obtained from the KT function, blue stars represent the experimental data points. Other solid circle lines show the simulated time spectra with various moment sizes. Best fit was obtained with $\mu_{Fe}$~$\sim$ 0.45~$\mu_{B}$. b) T-dependence of dc susceptibility of Fe$_{0.2}$Co$_{0.8}$Sn. The effective paramagnetic moment size, obtained from a Curie-Weiss fit (solid line), was found to be 0.81~$\mu_{B}$. The inset shows the T-dependence of the inverse of susceptibility. c) T-dependence of ac susceptibility, a pink arrow is a guide for the eye to show the frequency dependence of T$_{g}$. A Y-axis offset was used in data of 300~Hz, 4~Hz and 1~Hz for a better view of the frequency dependence.}
\label{Fe20_simu}
\end{figure}

To estimate the frozen static moment size of Fe, we performed simulations of the dipolar field by assuming two muon sites with a volume ratio of 60~$\%$: 40~$\%$ for $\mu_1$ and $\mu_2$ sites. The dipolar fields can be simulated for randomly allocated and oriented magnetic Fe atoms. Subsequently, we compared our simulations with the experimental ZF time spectra at 40 mK to estimate the effective moment size. Figure~\ref{Fe20_simu}(a) compares the simulated time spectra obtained for various Fe moment sizes with the experimental data and its fitting using Eq.\ref{M20Spinglass}. From this comparison, we obtain an effective moment size of $0.45(0.05)~\mu_B$. Note that the hyperfine field is not included in this argument, which gives a significant uncertainty to this estimate. Figure~\ref{Fe20_simu}(b) shows the results of our dc susceptibility measurements under an external field applied of 0.1~T. We fit the results with the Curie-Weiss law above 10~K, as shown by the solid blue line in Fig.~\ref{Fe20_simu}(b). The ordered moment size $\sim0.45~\mu_B$ is nearly half of the effective moment $0.81~\mu_B$ obtained from the magnetic susceptibility above $T_{g}$. Such reduction is commonly seen in many spin-glass systems. Figure~\ref{Fe20_simu}(c) shows the real part ($\chi '$) results of our ac susceptibility measurements at several frequencies with an excitation field of 0.3~mT after cooling the sample in the zero field. It exhibits the anomaly at $\sim$ 3.5~K for 1~Hz and was found to be frequency-dependent. The peak position shifts towards higher T with increasing frequency, consistent with a glass transition and $T_{g} \sim$ 3.5~K.

\section{DISCUSSIONS, CONCLUSIONS AND OUTLOOK}
\subsection{Intrinsic versus muon-specific anomalies in FeSn}
In the present $\mu$SR studies of FeSn, we found four notable features: i) the T~$\sim$~50~K anomaly of $1/T_{1}$ and $1/T_{2}$ in Figs.~\ref{ZF_fit} and \ref{T1}; ii) a signature of reorientation of local field direction towards c-axis direction at T $<$ 100~K and anomaly of ZF precession frequency at low temperatures in Figs.~\ref{ZF_fit}(a) and (b); iii) existence of the 'field cancelling muon site' with $\sim$ 40~$\%$ volume fraction; iv) anomaly of $1/T_{2}$ at T $\sim$ 250~K in Fig.~\ref{ZF_fit}(c) and very slow relaxation in nearly full amplitude in ZF at T $>$ 260~K in Fig.~\ref{ZF_wTFspec}(a) and the full amplitude oscillation at the applied wTF frequency at T = 270~K in Fig.~\ref{ZF_wTFspec}(d) and (f). Among them, i) is clearly due to intrinsic features of FeSn, as it is associated with the corresponding anomalies in ac and dc magnetic susceptibility shown in Figs.~\ref{res_sus}(b) and (c). Currently, the origin of this anomaly has not yet been identified. In an isostructural and ferromagnetic FeGe, a CDW transition is observed within the antiferromagnetically ordered state, as was identified by the associated anomalies in T-derivative of resistivity \cite{Teng2022} and structural x-ray measurements \cite{Miao2023}. In the present case of FeSn, no resistivity anomaly was found, as shown in Fig.~\ref{res_sus}(a) and no structural anomaly has been reported in neutron studies \cite{Xie2021}. Although the 'trimerization' symmetry-breaking on the Fe kagome plane was detected by a recent STM study of thin film FeSn at $\sim$ 4.5~K \cite{Zhang2023}, our STM results on cleaved bulk sample reveal no clear evidence for such rotational symmetry breaking as discussed in Sec.~\ref{STMsec}. To explore CDW, 'trimerization' and other possibilities, we are planning further studies of STM and high-resolution x-ray scattering on FeSn.

At this moment, it is not possible to unambiguously determine whether feature ii) is due to the intrinsic behaviour of FeSn or phenomena related to muon-specific features, such as muon site change, or both intrinsic and muon-specific phenomena. Here we would point out that very similar anomalies corresponding to the features i) and ii) have been observed in $\mu$SR measurements in planar antiferromagnetic systems CrSBr \cite{López-Paz2022}
having ferromagnetically correlated Cr moments in-plane, and antiferromagnetic REMn$_{6}$Sn$_{6}$\cite{MielkeIII2022} with Rare Earth elements RE = Y, and Tb, having ferromagnetically aligned Mn moments on kagome planes. It would be very interesting to perform further studies to seek if these common behaviours are due to the physics of layered metallic antiferromagnets with ferromagnetically correlated moments on kagome/triangular lattices and to identify if some muon specific feature is related or not to these observations.

The 'field cancelling site' iii) is the muon-specific phenomenon. In addition to the implication given by the 11~$\%$ Co-doped FeSn described in the present paper, we draw your attention to earlier Moessbauer effect studies on FeSn \cite{H_ggstr_m_1975, Kulshreshtha_1981, Ligenza_1971, Hartmann_1987}, which found responses from fully ordered volume below T$_{N}$ = 376~K, without any evidence of phase separation. The 250~K anomaly iv) is also a muon-specific result since no corresponding anomaly was seen in dc-and-ac magnetic susceptibility in Fig.~\ref{res_sus} and the Moessbauer effect signal in the ordered state around T $\sim$ 250~K. Therefore, we ascribe the 250~K anomaly to muon delocalization/diffusion between the high-field $\mu_1$ and field-cancelling sites $\mu_2$ shown in Fig.~\ref{muonsite}.

\subsection{Classical spin glass behavior in Fe$_{0.2}$Co$_{0.8}$Sn}\label{disc_B_spin_glass}

\begin{figure}
\centering\includegraphics[width=0.7\columnwidth]{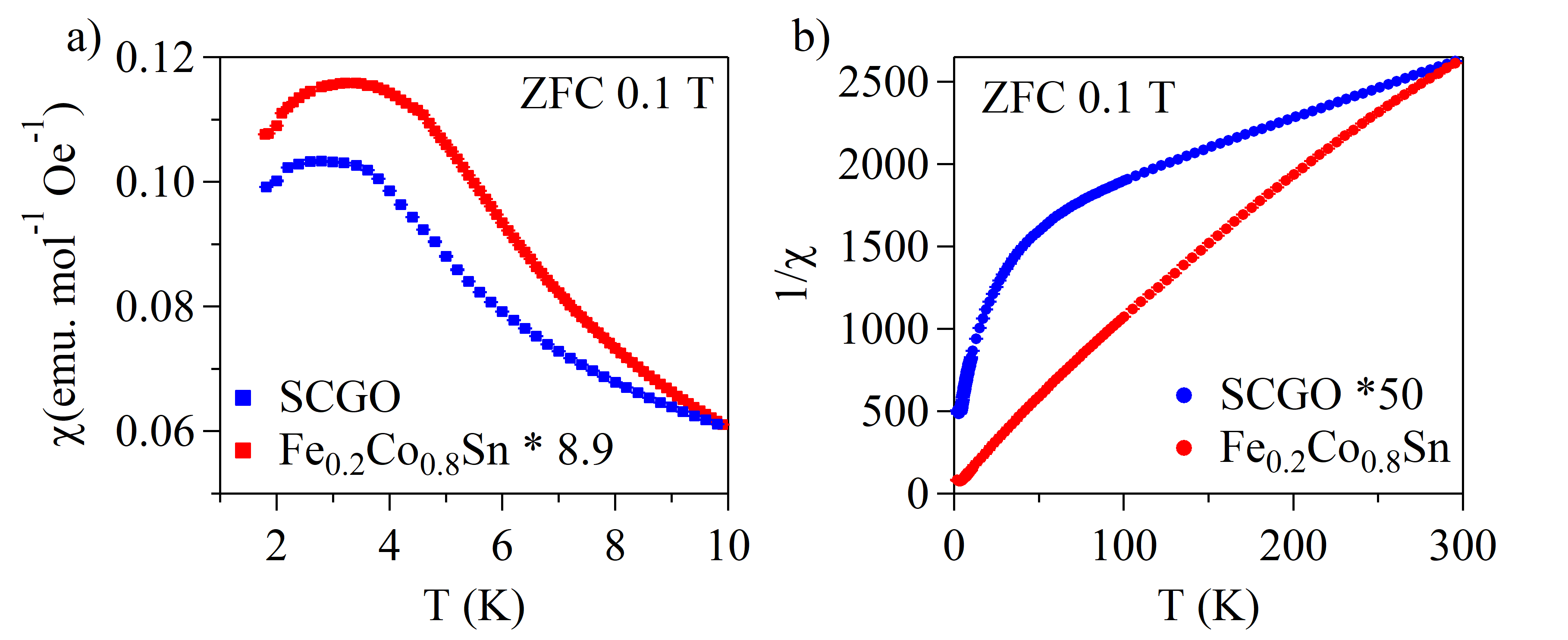}
\caption{Comparison of a) dc susceptibility and b) the inverse of dc susceptibility in ZFC mode under an external field of 1 kG. For better comparison, the dc susceptibility of Fe$_{0.2}$Co$_{0.8}$Sn was scaled by a factor of 8.9 in a), while the inverse of the susceptibility of SCGO was scaled by a factor of 50.}
\label{Fe20vsSCGO}
\end{figure}

In insulating kagome spin systems with antiferromagnetic nearest neighbor interactions, such as SrCr$_{8}$Ga$_{4}$O$_{19}$(SCGO) \cite{OBRADORS1988189}, KCr$_3$OH$_6$(SO$_4$)$_2$(Cr jarosite) \cite{Cr_jarosites}, and (Cu$_x$Zn$_{1-x}$)$_3$V$_2$O$_7$(OH)$_2\cdot2$H$_2$O(Volborthite) \cite{LAFONTAINE1990220}, geometrical frustrations lead to spin-glass like behaviours, while the dynamic relaxation observed by $\mu$SR \cite{Uemura1994, Keren1996, Keren2000, Fuyaka2003, Mendels2007, Bert_2004} exhibits nearly temperature-independent 'persistent quantum dynamics' below the $cusp$ temperature, associated with an unusual 'hard-to-decouple' line shapes. Similar behaviours were also observed in pyrochlore spin systems, such as NaCaNi$_2$F$_7$ and RE$_2$Ti$_2$O$_7$ \cite{Cai_2018, Lago_2005, Re_persistent}. On the other hand, dynamic spin fluctuations die away at low temperatures well below T$_{g}$ in typical dilute-alloy spin glasses CuMn, AuFe~\cite{Uemura1985}. Fig.~\ref{Fe20vsSCGO} compares the dc susceptibility for Fe$_{0.2}$Co$_{0.8}$Sn and SCGO with similar $T_{g} \sim$ 3.5 K in ZFC mode under the external field of 1 kG. The inverse of susceptibility demonstrates a much larger frustration index $|\theta_{CW}(150-300~K)|/T_{g} = 340 / 3.5$ $\sim$ 97 for SCGO as compared to 0.63 for Fe$_{0.2}$Co$_{0.8}$Sn. Note that $|\theta_{CW}|$ is also very small in AuFe and CuMn~\cite{Uemura1985}. The present study has identified that Fe$_{0.2}$Co$_{0.8}$Sn system behaves similarly to the dilute-alloy spin glasses. The internal field at the muon sites observed in Fe$_{0.2}$Co$_{0.8}$Sn is consistent in magnitude with the values expected for moments residing only on Fe atoms. In (Fe$_{1-x}$Co$_{x}$)Sn compounds, a spin-glass state appears when Fe concentration $(1-x)$ becomes smaller than the percolation threshold 0.5 of kagome plane, and the spin freezing temperature $T_{g}$ is proportional to the dilute Fe concentrations~\cite{Sales2021}. These features and similarities with CuMn and AuFe indicate that randomness and diluteness play primary roles in the origin of the spin glass behaviour of (Fe, Co)Sn. Although the kagome geometry leads to the flat band behaviours in band structures due to frustration in charge conductions in the underlying metallic lattice of CoSn, the present study demonstrated an example where dilute magnetic moments on the metallic kagome lattice exhibit the classical spin glass behaviour quite different from the spin-glass / spin-liquid behaviours often seen in insulating compounds with geometrical frustration of exchange interactions.

\section{ACKNOWLEDGMENTS}

We greatly appreciate the support of the personnel at TRIUMF during the $\mu$SR measurements. The work at Columbia was supported by the NSR grant DMR-2104661. The single crystal growth work at Rice was supported by US NSF DMR-2401084 (P.D.). The work at Princeton University was supported by the Institute for Quantum Matter, an Energy Frontier Research Center funded by the U.S. Department of Energy (DOE), Office of Science, Basic Energy Sciences under Award No. DE-SC0019331. The work in TRIUMF / UBC / SBQMI was financially supported in part by the Max Planck - UBC - UTokyo Center for Quantum Materials and the Canada First Research Excellence Fund, the Quantum Materials and Future Technologies Program, and the Discovery Grant from the Natural Sciences and Engineering Research Council of Canada (NSERC).

\bibliography{FeSn}

\end{document}